\documentclass{mn2e} \usepackage{graphicx} \usepackage{epsfig}
\usepackage{float} 
\usepackage{color} \usepackage{ulem}
\usepackage{longtable,lscape}
\def\starlight{\textsc{starlight}}                    
\newcommand{\hii}{\ifmmode [\rm{H}\,\textsc{ii}] \else [H~{\sc ii}]\fi}
\newcommand{\Ha}{\ifmmode {\rm H}\alpha \else H$\alpha$\fi}
\newcommand{\Hb}{\ifmmode {\rm H}\beta \else H$\beta$\fi}
\newcommand{\oiii}{\ifmmode [\rm{O}\,\textsc{iii}] \else [O~{\sc iii}]\fi}
\newcommand{\oii}{\ifmmode [\rm{O}\,\textsc{ii}] \else [O~{\sc ii}]\fi}
\newcommand{\oi}{\ifmmode [\rm{O}\,\textsc{i}] \else [O~{\sc i}]\fi}
\newcommand{\nii}{\ifmmode [\rm{N}\,\textsc{ii}] \else [N~{\sc ii}]\fi}
\newcommand{\sii}{\ifmmode [\rm{S}\,\textsc{ii}] \else [S~{\sc ii}]\fi}

\title[FRII radio galaxies in the SDSS]
      {FRII radio galaxies in the SDSS: Observational facts}

\author[Kozie\l]
{D. Kozie\l-Wierzbowska$^{1}$,
  G. Stasi\'nska$^{2}$\\
  $^{1}$Astronomical Observatory, Jagiellonian University,
ul. Orla 171, PL-30244 Krakow, Poland\\
  $^{2}$LUTH, Observatoire de Paris, CNRS, Universit\'e Paris
  Diderot; Place Jules Janssen 92190 Meudon, France\\
  }

\begin{document}

\maketitle

\begin{abstract} 
Starting from the Cambridge Catalogues of radio sources, we have created a sample of 401  FRII radio sources that have counterparts in the main galaxy sample of the 7th Data release of the Sloan Digital Sky Survey and analyse their radio and optical properties.

We find that the luminosity in the H$\alpha$ line -- which we argue gives a better measure of the total emission-line flux  than the widely used luminosity in \oiii\ -- is strongly correlated with the radio luminosity $P_{\rm 1.4GHz}$. We show that the absence of emission lines in about one third of our sample is likely due to a detection threshold and not to a lack of optical activity. 
We also find a very strong correlation between the values of $L_{\Ha}$ and $P_{\rm 1.4GHz}$ when scaled by $``M_{\rm BH}"$, an estimate of the black hole mass.

 We find that the properties of FRII galaxies are mainly driven by the Eddington parameter  $L_{\Ha}$/$``M_{\rm BH}"$ or, equivalently,  $P_{\rm 1.4GHz}$/$``M_{\rm BH}"$. Radio galaxies with hot spots are found among the ones with the highest values of $P_{\rm 1.4GHz}$/$``M_{\rm BH}"$. 

Compared to classical AGN hosts in the main galaxy sample of the SDSS, our FRII galaxies show a  larger proportion of objects with very hard ionizing radiation field and large ionization parameter. A few objects are, on the contrary, ionized  by a softer radiation field. Two of them have  double-peaked emission lines and deserve more attention. 

We find that the black hole masses and stellar masses in FRII galaxies are very closely related:  $``M_{\rm BH}"$ $\propto$ $M_{*}^{1.13}$ with very little scatter. A comparison sample of line-less galaxies in the SDSS follows exactly the same relation, although the masses are, on average, smaller. This suggests that the FRII radio phenomenon occurs in normal elliptical galaxies, preferentially in the most massive ones. Although most  FRII galaxies are old, some  contain traces of young stellar populations. Such young populations are not seen in normal line-less galaxies, suggesting that the radio (and optical) activity in some FRII galaxies may be triggered by recent star formation. The $``M_{\rm BH}"$ -- $M_{*}$ relation in a comparison sample of radio-quiet AGN hosts from the SDSS is very different, suggesting that galaxies which are still forming stars are also still building their central black holes. 

Globally, our study indicates that, while radio and optical activity are strongly related in FRII galaxies, the features of the optical activity in FRIIs are distinct from those of the bulk of radio-quiet active galaxies. 
 
An appendix gives the radio maps of our FRII galaxies, superimposed on the SDSS images, and the parameters derived for our analysis that were not publicly available.

  \end{abstract}



\section{Introduction}
\label{sec:Introduction}

According to the Collins English Dictionary,  the term "radio galaxy" (RG) refers to  a galaxy that is a strong emitter of radio waves.   However, only sources powered by 
 accretion onto a super massive black hole can produce the extended structures 
called radio jets and lobes, and we will use the term radio galaxy in reference
only to those sources. 
Accretion unto a massive black hole is what is considered to be the energy source of active galactic nuclei (AGN). It
manifests itself in different ways and with different strengths in the entire  electromagnetic spectrum. 
Most active galactic nuclei in the Sloan Digital Sky Survey (SDSS, York et al. 2000) are not radio active.
Conversely, not all classical radio galaxies with extended radio lobes have 
emission lines in their optical spectra (Hine \& Longair, 1979),  suggesting  that 
optical and radio activity are  not necessarily concommittent  (Best et al. 2005b).

 Radio galaxies can be divided into two classes according to the morphology of 
their radio structure  (Fanaroff \& Riley, 1974). FRI radio 
galaxies are core-dominated   sources with radio jets fading 
and dissipating on a short distance from center, while FRII radio 
galaxies are edge-brightened  sources with highly collimated 
jets. The sizes of both types of radio galaxies range from a few kiloparsecs for compact 
steep spectrum   sources to a few megaparsecs  
for giant radio galaxies.  FRII radio galaxies tend to be more luminous than FRI ones.  Fanaroff \& Riley suggested a dividing 
luminosity of $L_{178MHz} \sim$ 2.5 $\times$ 10$^{26}$W Hz$^{-1}$, but as shown 
by Ledlow \& Owen (1996) the dividing luminosity between FRI and FRII radio sources 
is a function of the host galaxy optical luminosity.

With the availability of observational data from large radio 
and optical surveys came the possibility to examine the optical properties of large samples of radio galaxies.
 Best et al. (2005a) cross-identified radio sources from the NVSS (Condon et al. 1998) and FIRST (Becker et al. 1995) radio surveys with the main galaxy sample of the second data release of the SDSS (Abazajian et al. 2004). The resulting sample, which is not morphology-specific is likely dominated by FRI or compact radio-sources, as can be judged by the low radio luminosities of the vast majority of their sample.

 Using this (or a similar) sample Kauffmann, Heckman \& Best (2008) have argued 
that radio emission is likely due to the accretion of \textit{hot} gas by massive black holes 
central to galaxies in a dense environment, while the optical AGN phenomenon, strongly favoured 
by the presence of a young stellar population, is likely due to the accretion of\textit{ cold}
gas.  However, their sample lacks the brightest radio sources, the FRII ones, both because they are rare  in the local Universe and because the very extended angular sizes of these sources does not allow easy cross-identification 
with optical galaxies.

FRII radio galaxies constitute a much better defined class 
than FRI radio galaxies in terms of radio morphology 
(Fanaroff \& Riley, 1974). Besides, because of the diversities of their optical properties, 
they constitute a prime target for  understanding the relation between optical and radio activity.  

We have assembled a sample of  morphologically selected FRII galaxies with optical counterparts available in the SDSS. 
Here, we present the result of our study comparing the radio and optical properties of FRII 
radio galaxies, including giant radio galaxies, in order to get more insight into the activity phenomenon in galaxies. Our data 
sample is, by necessity, much smaller than the sample used by Kauffmann et al. (2008), since FRII galaxies are much less common than FRI galaxies at low redshifts. Due to the selection process (see next section), our sample  does not 
allow us to tackle such issues as luminosity distribution functions, which are examined by 
Kauffmann \& Heckman  (2009) using the same sample as Kauffmann et al. (2008). But we can look for the presence 
of correlations that might improve our understanding of radio loud AGN.

The organization of the paper is as follows:
In Section 2 we provide a brief description of the sample selection and data processing.
In Section 3 we analyze the relation between the radio power of FRII sources and the strength of their optical activity. In Section 4 we focus on the special class of FRII galaxies that show hot spots. In Section 5 we discuss the emission line properties of FRII galaxies. In Section 6 we put FRII galaxies in the context of other groups of galaxies: line-less galaxies and radio-quiet AGN hosts. The main results of our investigation are summarized in Section 7.

Throughout this paper we assume a $\Lambda$ Cold Dark Matter
cosmology with $H_{0}=71$km\,s$^{-1}$Mpc$^{-1}$, $\Omega_{\rm m}=0.27$, and
$\Omega _{\Lambda}=0.73$ (Spergel et al. 2003).
 

\section{The data}
\label{sec:data}

\subsection{The sample}
\label{sec:sample}

The main aim of our selection  was to obtain a sample of FRII radio galaxies 
with a large range of radio powers and sizes in order to study the relation between their radio properties with the optical properties of their 
associated galaxies.   Unfortunately,  an automatic cross-correlation of radio 
and optical catalogues misses radio sources with large angular size,
without radio core or with radio flux at the catalogue limit. On the other hand the luminosity profiles of radio structures are quite 
complex and the algorithms of automatic selection from radio maps are not 
sophisticated enough to recognize   radio structures of different 
luminosity profiles. Thus the method used here is a combination of automatic 
and manual selection from catalogues and radio maps  and we had to restrict ourselves to a tractable number of radio maps to examine.

For this purpose, we limited our search to radio sources present in the  Cambridge Catalogues of Radio Sources: 3C (Edge et al. 1959; Bennett 1962), 
4C (Pilkington \& Scott 1965; Gower, Scott \& Wills 1967), 
5C (Pearson 1975; Pearson \& Kus 1978; Benn et al. 1982; Benn \& Kenderdine 1991; Benn 1995), 
6C (Baldwin et al. 1985; Hales, Baldwin \& Warner 1988; Hales et al. 1990, 1991, 1993; Hales, Baldwin \& Warner 1993), 
7C (Hales et al 2007), 8C (Rees 1990; Hales et al. 1995) and 9C (Waldram et al. 2003) 
and using the SDSS CrossID we crossidentified them with the sample of 926246 galaxies from the SDSS DR7 main galaxy sample (Abazajian 2009), keeping only those radio sources whose  optical spectra are available in the SDSS. 
Taking into account the sometimes large positional uncertainties in the Cambridge Radio Catalogues we 
adopted 0.2 arcmin maximum distance between radio core 
and optical galaxy. Next, we excluded all objects classified in the SDSS 
as quasars. We postpone the study of the quasar sample of the SDSS DR7
(which also contains the Seyfert 1 galaxies) to a next paper.
The Cambridge Radio Catalogues were based on radio maps made at different radio frequencies
and with different resolution, thus morphological classification based on these maps would give incoherent results not easy to compare. 
Therefore, in a third step we inspected the NVSS and FIRST (if available) radio maps
of all the remaining sources (almost 2000). These maps were made at the same frequency (1.4 GHz),
but with different resolutions, what facilitates identification of the most common 
features in FRII radio galaxies which are  extended radio lobes (NVSS), compact cores and hot spots (FIRST).
Thereby we were able to remove FRI objects as well as sources  
with disturbed radio morphology or with angular size too small 
to determine the morphology (mostly sources with angular diameter $\theta$ $<$ 20 arcsec). 
In this step we also excluded misidentified sources.

The pre-selected sample contained only few sources with radio structures
larger than 700 kpc, known as giant radio galaxies. However, there is a significant number of 
giant radio galaxies that can be included into our analysis. Therefore, we 
considered the known ones (using the lists of Janda 2006 and Machalski et al. 2007), and looked for 
their optical counterparts in the SDSS DR7 spectroscopic catalogue.  
This allowed us to include all  the giant radio galaxies with available spectroscopic data 
into our final sample.  

We  thus  obtained a sample of 401 FRII radio galaxies, 
out of which  23 have  diameters larger than 700 kpc.  
Note that our sample is not complete in any sense and is not adequate to study luminosity functions. We used radio catalogues that were based on
 radio observations at different frequencies and made with different flux limits.
But our sample does cover a wide range of radio powers and sizes. 

The optical data (magnitudes and spectroscopy) come from the SDSS. The spectra were taken with 3 arcsec diameter fibers and cover a wavelength 
range of 3800--9200 \AA\ with a mean spectral resolution of 1800. We use 
the data as given in the seventh data release.
 
Obviously, inferences derived from 
 a given  sample are not necessarily valid for another one. Many studies (e.g. Best. et al. 2005a,b; Kauffmann et al. 2008;  Smol\v ci\'c et al. 2009;
 Smol\v ci\'c 2009) include different morphological types of radio galaxies without distinguishing among them. Those samples, however, go to much lower 
 radio luminosities than ours, and are thus complementary in some respect. The  Smol\v ci\'c et al. (2009) sample is extracted from the \textsc{VLA-COSMOS} survey 
 (Schinnerer et al. 2007), and has the advantage of the existence of many ancillary data at all wavelengths. Other samples (e.g. Rawlings et al. 1989, Zirbel \& Baum 1995, 
 Owen \& Ledlow 1994), while focusing on FRII types or explicitly distinguishing them from FRI ones, have only limited information on the properties that 
 can be derived from optical data, such as black hole masses or accurate emission line fluxes in the case of lines with small equivalent widths. 
The sample of Buttiglione et al. (2009, 2010) is extracted from the 3CR radio catalogue only, imposing a radio flux limit higher than ours. 
It distinguishes between FRII, FRI and so-called compact sources. The optical data come mostly from their own observations. But the analysis of the stellar 
continuum is not as advanced as ours (see Sect. 2.2), and parameters such as galaxy masses or black hole masses were not obtained. 
Thus our sample, although restricted to the Cambridge Catalogues of Radio Sources,
is the only one that allows the exploration of the optical properties of FRII galaxies 
(including giants),  comparing radio properties with galaxy masses or accretion rates on the central black hole. The redshifts in our sample range from 0.045 to 0.6.

\subsection{Data processing}
\label{sec:optical}

The 1.4 GHz radio luminosities, $P_{\rm 1.4GHz}$, of all the sources were calculated from 
the total flux density at this frequency  obtained as a sum of fluxes of all components fitted to the each 
source and listed in the NVSS catalogue. 

The angular sizes of the sources, defined as the distances between 
the hot spots or between the most distant bright structures in both lobes, as it is in the case of relic or FRI/II radio sources, 
were estimated manually either from the FIRST maps if available, or from the NVSS maps using the Aladin Sky Atlas. 
The manual method of deriving the angular size involves an error in this quantity of about 10\%.
The angular sizes were used to determine the linear 
(projected) diameters of the sources, $D$. In our sample, these  extend 
from 15 kpc to 2080 kpc.

The optical parameters of the sample galaxies, i.e. their stellar masses and ages 
as well as the emission line fluxes 
are taken from the \starlight\ database\footnote{see http://www.starlight.ufsc.br} 
(Cid Fernandes et al. 2009).  
\starlight\ (Cid Fernandes et al. 2005, see also Mateus et al. 2006) recovers 
the stellar population of a galaxy by fitting a pixel-by-pixel model to the spectral 
continuum (excluding narrow windows where emission lines are expected as well as bad pixels). 
The model is a linear combination of 150 simple stellar populations templates 
 with ages 1 Myr $\le t_\star \le$ 18 Gyr, and metallicities $0.005 \le Z/Z_\odot \le 2.5$.
 In this paper we use parameters that were calculated in the same way as in Cid Fernandes et al. (2010), i.e. using 
Bruzual \& Charlot (2003) evolutionary stellar population models,  with the
STELIB library (Le Borgne et al. 2003), ``Padova 1994'' tracks
(Bertelli et al. 1994)  and Chabrier (2003) initial mass
function.  Emission lines fluxes are measured by Gaussian fitting in the residual spectra,
which reduces the contamination by stellar absorption features. We checked, by visual inspection, 
all the SDSS spectra and corrected the line intensities in the very rare cases where 
the automatic procedures led to spurious results.

\begin{figure}
\centering
\includegraphics[width=8cm,angle=0]{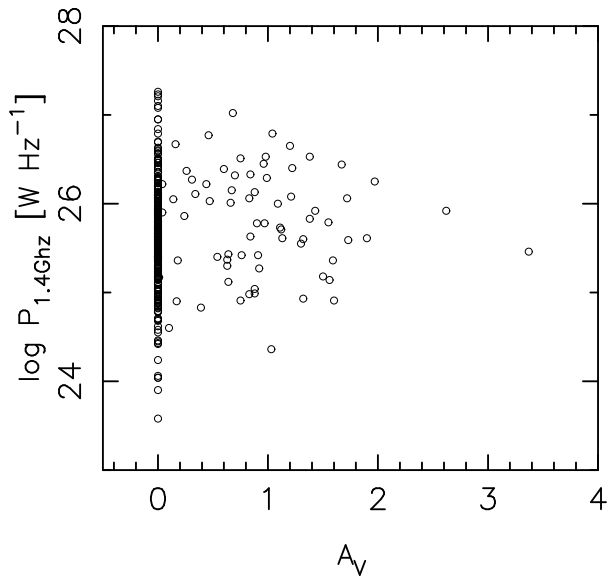}
\caption{The relation between the 1.4 GHz radio luminosity, $P_{\rm 1.4GHz}$, and $A_{\rm V}$ for our sample of FRII radio galaxies.}
\label{fig:f1}
\end{figure}

To correct the emission lines for extinction, 
one usually forces the observed H$\alpha$/H$\beta$ ratio to the theoretical
case B recombination value of 2.9. Unfortunately, in our sample, there are
many galaxies which do not have both
H$\alpha$ and H$\beta$ fluxes measured with sufficient accuracy. In a
preliminary step, we have computed the visual extinction, $A_{\rm V}$, using the Fitzpatrick (1999) extinction law parametrized with
$R_{\rm V}$ =3.1 for all the objects in our sample having $S/N > 3$ in both
H$\alpha$ and H$\beta$. Figure \ref{fig:f1} shows $P_{\rm 1.4GHz}$ vs
$A_{\rm V}$. Objects for which $A_{\rm V}$ could
not be evaluated are represented  at an abscissa of 0.  
We see that there is no  relation between the 
 $A_{\rm V}$ and $P_{\rm
1.4GHz}$, in particular no indication of a tendency for $A_{\rm V}$ to increase with decreasing $P_{\rm
1.4GHz}$. It is therefore likely that the extinction is in fact
between $\sim$ 0.5 and $\sim$ 1.2 for most of our galaxies for which we could not
obtain it. 
Therefore, correcting the line intensities for extinction in some of
the objects and not  in others  is not better justified than performing no extinction correction at all.  In this paper we use only uncorrected line
intensities  (except in line ratio diagrams where the dereddening could be applied to all the objects in the plot)\footnote{As a matter of fact, we also repeated the entire
analysis by applying the extinction correction where we could, some of the
plots presented here as well as their corresponding regression lines are
slightly changed, but nothing important on statistical grounds}.

 As mentioned above, the stellar masses used in this paper, $M_*$, were taken from the \starlight\ database. They were obtained from the stellar 
masses corresponding to the light inside the fiber  by correcting for aperture 
affect assuming that the mass-to-light ratio outside the fiber is the same 
as inside and scaling the fiber masses by the ratio between 
total (from the photometric data base) and fiber $z$-band luminosities. This correction is smaller 
than a factor of 2 in a large portion of our sample, but can amount to factors of up to 8. 
On the other hand, we do not correct the emission line luminosities for aperture effects, 
since the emission lines are expected to be emitted in the inner regions of the galaxies 
associated with the radio sources.

In this paper, we will often refer to the black hole masses of the galaxies, estimated from the observed stellar velocity dispersion given by the SDSS, $\sigma_{*}$,  using the popular relation  by Tremaine et al. (2002): 
\begin{equation}
\label{eq: MBH}
{\rm log} M_{\rm BH} = 8.13 + 4.02\, {\rm log} (\sigma_{*} / 200\,{\rm km\,s^{-1}}).
\end{equation}

The interpretation of such a formula is a challenge, since it seems to indicate a causal relation between the masses of two types of objects (galaxy bulges and black holes) that differ in size by 3--4 orders of magnitude. 
In particular, its range of validity is not clearly established. 
In the following, we will use the notation $``M_{\rm BH}"$ to refer to the ``black hole mass'' as derived from Eq. \ref{eq: MBH} 
to remember that it does not necessarily represent the true mass of the central black hole, but is merely 
an expression derived from the measured velocity dispersion. 

We consider $M_*$ values only for objects with sufficiently good spectra (in practice we impose 
a S/N $>10$ in the continuum). The $``M_{\rm BH}"$ values will be considered only for those objects 
which, in addition, have $\sigma_{*} > 60$\,km\,s$^{-1}$. 
In such a way, the values of $M_*$ and $``M_{\rm BH}"$ that we use will not be strongly affected by observational errors.


\section{The relation between the radio power and optical activity of FRII sources}
\label{sec:radio-opt}

\begin{figure}
\centering
\includegraphics[width=8cm,angle=0]{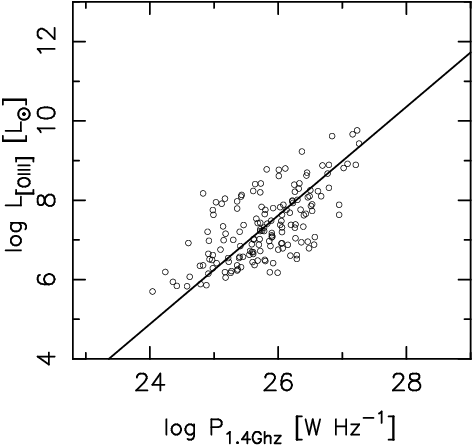}
\caption{$L_{\oiii}$  versus $P_{\rm 1.4GHz}$, for FRII galaxies having $S/N > 3$ in the \oiii\ line. The straight line is the bissector regression line.}
\label{fig:f2}
\end{figure}

Rawlings et al. (1989) were the first to note the existence of an intrinsic correlation between the radio power and $L_{\oiii}$ in a sample of 39 FRII galaxies. They take it as evidence for a physical coupling of the processes that supply energy to the emission line region and to the extended radio source. Our sample  confirms this trend and shows a very strong correlation, as seen in Fig. \ref{fig:f2}. The Pearson correlation coefficient is $R$=0.66 for a sample of 160 objects having the \oiii\ line measured with a signal-to-noise ratio ($S/N$) larger than  3. Using the regression package of Akritas \& Bershady (1996), the  bissector regression line, assuming a typical error of 0.2 dex in all the observables, is found to be: 
\begin{eqnarray}
{\rm log} L_{\oiii} = (1.37 \pm 0.07)  \times {\rm log} P_{\rm 1.4GHz}  \nonumber\\
+ (-28.0 \pm 1.7).
\end{eqnarray}

 Note that, if $L_{\oiii}$ is taken to be a measure of the AGN luminosity in radio galaxies, 
 one should in principle worry about the possible contribution of \hii\ regions.  
Kauffmann \& Heckman (2009) have proposed a rough method to 
correct for this effect by taking into account the galaxy distance from the star forming branch in the BPT diagram.
 This, indeed should improve the estimate of the AGN luminosity 
of radio galaxies that experience star formation. In our sample of FRII galaxies, though, 
there is no star formation occuring presently, as argued later in this paper, therefore such a correction is not needed.
\begin{figure}
\centering
\includegraphics[width=8cm,angle=0]{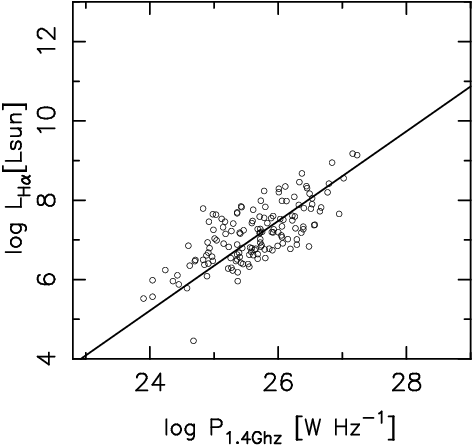}
\caption{$L_{\Ha}$  versus $P_{\rm 1.4GHz}$, for FRII galaxies having $S/N > 3$ in the \Ha\ line. The straight line is the bissector regression line.}
\label{fig:f3}
\end{figure}

The reason for using  $L_{\oiii}$ as a way to estimate the total energy emitted in the lines is that \oiii\ is often the strongest line in optical spectra. As a matter of fact,  it is much better to use the luminosity of \Ha,  $L_{\Ha}$. Even if \Ha\ may be weaker than \oiii, the fact that its intensity is proportional to  Ly$\alpha$,  which is 8 times stronger, makes of it a far more reliable indicator of the total energy emitted in the lines. Of course, the bolometric correction factor will not be 
the same for $L_{\oiii}$ and $L_{\Ha}$. A further argument for preferring $L_{\Ha}$ to $L_{\oiii}$ is that its value is independent of the ionization state, contrary to $L_{\oiii}$. Fig. \ref{fig:f3}, shows $L_{\Ha}$ as a function of $P_{\rm 1.4GHz}$, for the 145 FRII galaxies having $S/N > 3$ in the \Ha\ line. As expected, the correlation is  better: $R$=0.72. The bissector regression line is: 
\begin{eqnarray}
{\rm log} L_{\Ha} = (1.13 \pm 0.07)  \times {\rm log} P_{\rm 1.4GHz} +   \nonumber\\
(-21.9 \pm 1.8).
\end{eqnarray}

Thus, both $L_{\oiii}$ and  $L_{\Ha}$ (when available) indicate a strong correlation between AGN activity and radio power of these extended radio sources. While qualitatively, this result is in agreement with the one found by Zirbel \& Baum (1995) who used \Ha + \nii, there is a  significant difference. Zirbel \& Baum (1995) found that, for the FRII radio galaxies they considered, the exponent of the relation between  $L_{\Ha + \nii}$ and radio power is $0.75\pm 0.09$, while we find that the exponent of the relation between $L_{\Ha}$ and radio power is as large as $1.13 \pm 0.07$. We checked that this difference is not due to the \nii\ line, whose contribution might a priori change systematically with luminosity. 
As a matter of fact, it seems that the difference is mainly due to the way 
the regression line is obtained.

Buttiglione et al. (2010) found that high excitation radio galaxies (HEGs) follow a slightly 
different $L_{\oiii}$ vs $P_{\rm 1.4GHz}$ relation than low excitation radio galaxies (LEGs). 
We believe that this is simply the consequence of the fact that $L_{\oiii}$ is strongly dependent of the ionization state and 
is therefore a biased estimator of the total AGN energy emitted in the lines. Note that $L_{\Ha}$, 
in addition to being a good estimator of the energy emitted in the lines, is also 
an\textit{ exact } estimator of the total number of ionizing photons emitted by the AGN 
(if all of them are absorbed by the gas).

\begin{figure}
\centering
\includegraphics[width=8cm,angle=0]{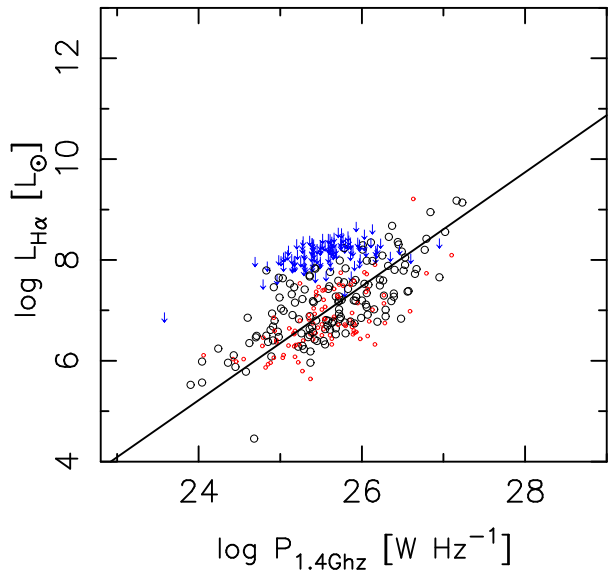}
\caption{$L_{\Ha}$  versus $P_{\rm 1.4GHz}$  for FRII galaxies of our sample. Black circles correspond to objects having $S/N > 3$ in the \Ha\ line, small (red in the electronic version) circles to objects where the  \Ha\ line is detected with a lower S/N.  The grey (blue in the electronic version) arrows correspond to objects where \Ha\ is not detected although not redshifted out of the SDSS wavelength range, and represents the lower limit in \Ha\ flux that could be detected given the S/N of the spectrum in the continuum. The straight line is the same as in Fig. \ref{fig:f3}.}
\label{fig:f5}
\end{figure}

\begin{figure}
\centering
\includegraphics[width=8cm,angle=0]{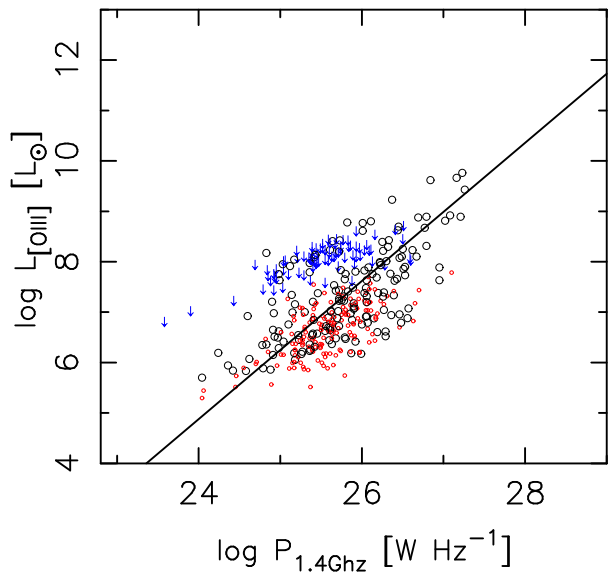}
\caption{$L_{\oiii}$  versus $P_{\rm 1.4GHz}$  for FRII galaxies of our sample. The presentation is analogous to that in Fig. \ref{fig:f5}. The straight line is the same as in Fig. \ref{fig:f2}.}
\label{fig:f8}
\end{figure}

Figures \ref{fig:f2} and \ref{fig:f3} were constructed by using only data for which the relevant emission lines have a S/N $>3$. However, even including data with much worse S/N, as shown in Fig. \ref{fig:f5} with small (red in the electronic version) circles points for the case of $L_{\Ha}$, the regression lines remain the same and the dispersion is not increased. This is because the correlation is so strong over a range of several decades in optical and radio luminosities while the measurement of a line intensity cannot be wrong by a factor more than $\sim$ 2 when it is detected, which is negligible with respect to the range of luminosities encountered. 

The fact that such a strong correlation exists for the 298 FRII galaxies with the \oiii\ line 
detected and the 205 ones with the \Ha\ line detected, in a total sample of 401 radio galaxies 
(of which 45 have redshifts larger than 0.4, which shifts the \Ha\ line out of the SDSS wavelength 
range) raises the question of whether this is a universal relation for FRII galaxies. Using the S/N 
ratio in the continuum of the objects where those lines have not been detected 
and assuming a typical emission-line width of  10\AA, we have estimated the minimum detectable flux in these 
lines for each object. These numbers, converted into luminosities, are plotted as grey (blue in the electronic version) arrows 
in Fig. \ref{fig:f5}. One can see that the objects with undetected lines in \Ha\ could well follow 
exactly the same trend as the ones with detected \Ha. A similar figure for \oiii\ (Figure 
\ref{fig:f8}) gives the same result. Note that this is at variance with the finding by 
Buttiglione et al. (2010) that, in their sample (which however does not contain only FRII galaxies), 
for objects with undetected \oiii\ the upper limits on \oiii\ luminosities are well below 
the prediction of the correlation between emission line and radio power.  Thus, we find 
no evidence for the existence of line-less FRII radio galaxies. 

\begin{figure}
\centering
\includegraphics[width=8cm,angle=0]{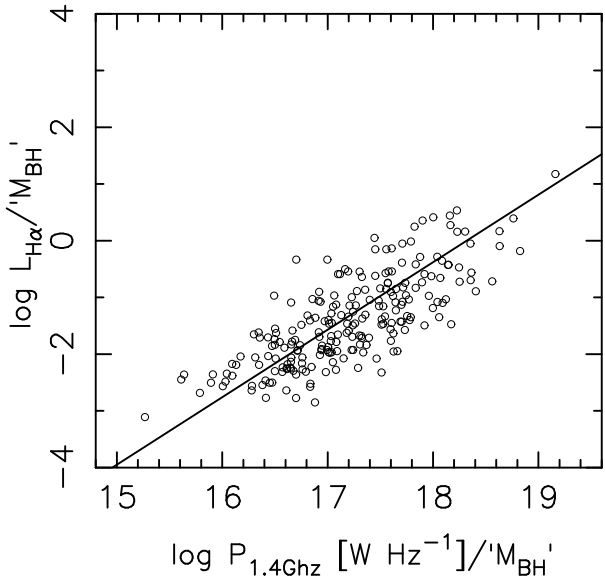}
\caption{The relation between $L_{\Ha}$/$``M_{\rm BH}"$ and  $P_{\rm 1.4GHz}$/$``M_{\rm BH}"$  for FRII galaxies having $S/N > 3$ in the \Ha\ line. The straight line is the bissector regression line.}
\label{fig:f4}
\end{figure}

In their sample of radio galaxies (of all types), Kauffmann et al. (2008) did not find a strong correlation between $L_{\oiii}$ or$L_{\Ha}$ and $P_{\rm 1.4GHz}$, but when scaling the emission-line and radio luminosities by the black hole mass, they do find  some correlation between normalized radio power and accretion rate.
 
 In Fig. \ref{fig:f4}, we show 
 $L_{\Ha}$/$``M_{\rm BH}"$ vs $P_{\rm 1.4GHz}$./$``M_{\rm BH}"$ for our sample 
 of FRII radio galaxies having  $S/N > 3$ in the \Ha\ line.  The correlation coefficient is $R$=0.77. To our knowledge, this is the first time that such a strong correlation 
 is shown to exist between  $L_{\Ha}$/$``M_{\rm BH}"$ and $P_{\rm 1.4GHz}$./$``M_{\rm BH}"$ in powerful radio galaxies.  
The bissector regression line is: 

\begin{eqnarray}
{\rm log} L_{\Ha}/``M_{\rm BH}" = (1.19 \pm 0.07)  \times {\rm log} P_{\rm 1.4GHz}/``M_{\rm BH}" +   \nonumber\\
(-21.8 \pm 1.3).
\end{eqnarray}

\begin{figure}
\centering
\includegraphics[width=8cm,angle=0]{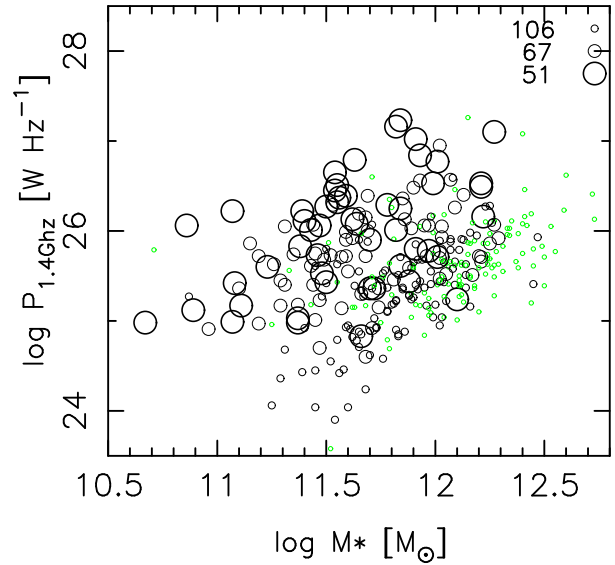}
\caption{The relation between: $P_{\rm 1.4GHz}$ and the stellar masses  $M_*$ of the FRII galaxies. The symbol sizes correspond to values of the ``Eddington parameter'' $L_{\Ha}$/$``M_{\rm BH}"$. Large circles: log $L_{\Ha}$/$``M_{\rm BH}"$ $>-0.8$ ; intermediate size circles: log $L_{\Ha}$/$``M_{\rm BH}"$ between  $-1.5$ and $-0.8$; small black circles: log $L_{\Ha}$/$``M_{\rm BH}"$ $< -1.5$; very small grey (green in the electronic version) circles: \Ha\ not detected. }
\label{fig:f6}
\end{figure}

\begin{figure}
\centering
\includegraphics[width=8cm,angle=0]{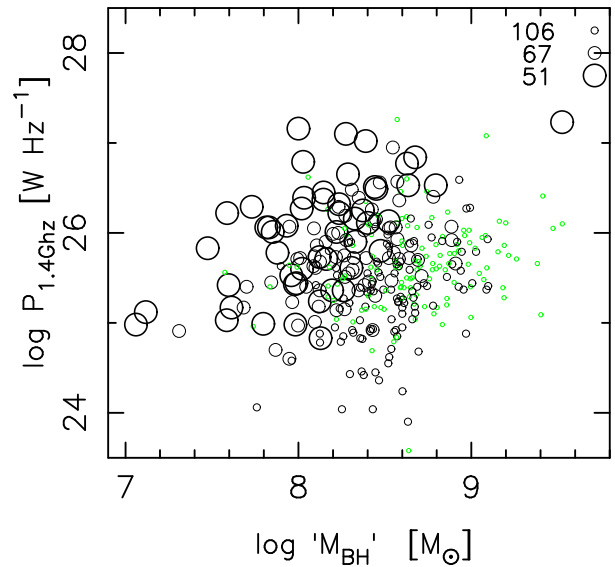}
\caption{The relation between: $P_{\rm 1.4GHz}$ and $``M_{\rm BH}"$ for the FRII galaxies. Same symbols as in Fig. \ref{fig:f6}.}
\label{fig:f7}
\end{figure}

In Fig.  \ref{fig:f6} we investigate how the value of the ratio  $L_{\Ha}/``M_{\rm BH}"$ (from now on referred to as the Eddington parameter, by similarity with Kauffmann \& Heckman 2009)  is distributed in the  ($P_{\rm 1.4GHz}$, $M_*$) plane. The sizes of the black circles correspond to the value of  $L_{\Ha}/``M_{\rm BH}"$, as indicated in the figure caption. The small grey (green in the electronic version) circles represent objects for which \Ha\ is not observed. Roughly, the circle sizes decrease perpendicularly to the weak trend between $P_{\rm 1.4GHz}$ and $M_*$ (the Pearson correlation coefficient between $P_{\rm 1.4GHz}$ and $M_*$ is $R$=0.31). This means that the most powerful radio galaxies are also the ones with the largest Eddington parameter, for a given galaxy mass, and that, for a given radio power, the Eddington parameter increases as the galaxy mass decreases. Note that the lower envelope of the points in this diagram increases with increasing  $M_*$. This is a transcription, in the  ($P_{\rm 1.4GHz}$, $M_*$) plane, of the fact that the FRII/FRI transition occurs in a narrow zone of the radio luminosity - optical luminosity diagram as shown by Owen \& Ledlow (1994).

In the  ($P_{\rm 1.4GHz}$,$``M_{\rm BH}"$) plane, as seen in Fig. \ref{fig:f7}, the distribution of $L_{\Ha}/``M_{\rm BH}"$  with respect to the axes is similar to that seen in Fig.  \ref{fig:f6}, i.e. perpendicular to the first diagonal (as implied by the strong correlation between $P_{\rm 1.4GHz}$ and  $L_{\Ha}$). But the radio power and $``M_{\rm BH}"$ are only loosely correlated ($R$=0.15).
 

\section{Properties of FRII radio sources with hot spots}
\label{sec:morpho_gal}

Fanaroff \& Riley (1974) used the ratio of the distance between the brightest 
regions in lobes placed on opposite sides of a central galaxy to the total 
source size  as a criterion to classify extended radio sources.
The regions of the highest brightness at the end of lobes are places where 
the relativistic jets emanating from active galactic nuclei interact
with the environment, and are called hot spots (the detailed definition 
of hot spots can be found in Hardcastle et al., 1998). Not all radio sources show hot spots. For example, in the case of radio relicts, i.e. sources where the central activity has 
already stopped, hot spots are not present (Kaiser et al. 2000, 
Marecki \& Swoboda, 2011).
One can thus expect that the presence of hot spots is related with some properties of the AGN and its environment. 

We have searched our sample of FRII radio  sources for the presence of hot spots. We were able to do this only for sources 
with available FIRST radio maps (whose resolution is much better  than that of NVSS 
maps) and with angular size large enough to separate the bright, compact components 
from the bright lobes. Of  211   sources with suitable radio maps and large enough angular sizes, 51  were found to have prominent hot spots.

Owen and Laing (1989) proposed an alternative way of classifying radio sources in
terms of morphology. They introduced 3 classes of radio sources:
Twin Jet, Classical Double and Fat Double sources. 
Twin Jet galaxies fit into the FRI type, Classical Doubles are galaxies with compact hot spots 
and elongated lobes and fit into the FRII class. Fat Double galaxies with diffuse 
lobes and bright outer rims can be included into FRII or FRI/II classes.
Our FRII radio galaxies with hot spots are thus genuine Classical Doubles.
Owen and Laing showed, in their Fig. 6, that Classical Double sources are more luminous 
in radio  but less luminous in the optical  than Twin Jet and Fat Double radio galaxies. 

\begin{figure}
\centering
\includegraphics[width=8cm,angle=0]{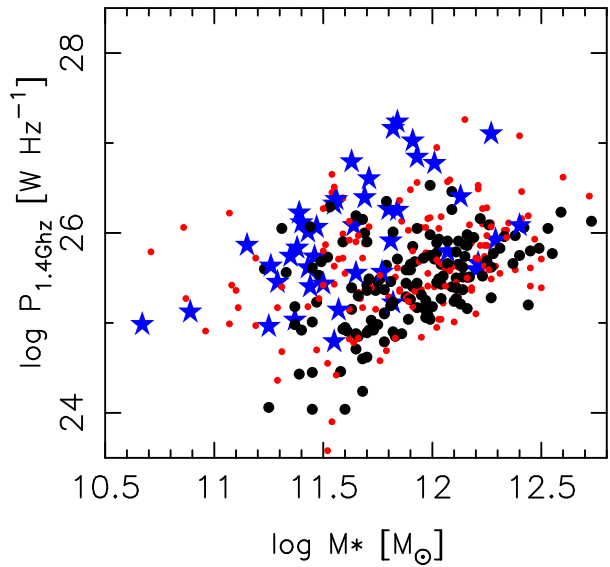}
\caption{The distribution of the FRII radio sources  with hot spots in the ($P_{\rm 1.4GHz}$, $M_*$) plane. Stars: sources with hot spots; Big dots: sources without hot spots. Small dots: Sources for which the classification could not be made. }
\label{fig:f9}
\end{figure}

\begin{figure}
\centering
\includegraphics[width=8cm,angle=0]{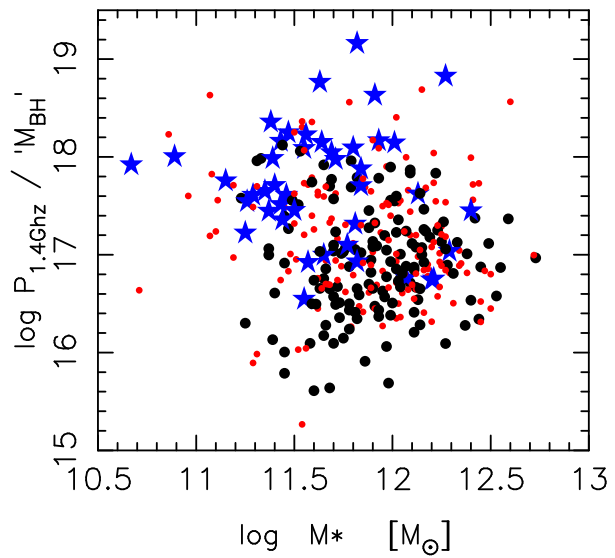}
\caption{The distribution of the FRII radio sources  with hot spots in the ($P_{\rm 1.4GHz}$/$``M_{\rm BH}"$, $M_*$) plane. The symbols are the same as in Fig. \ref{fig:f9}.}
\label{fig:f10}
\end{figure}

We show, in Fig.  \ref{fig:f9}, how radio galaxies with and without 
hot spots are located in the $P_{\rm 1.4GHz}$ vs $M_*$ diagram. Sources with
prominent hot spots  are represented by stars. 
 Sources which definitely do not show any hot spot are represented by big dots. Sources for which we could not do the classification are represented by small dots.
 As can be seen in the figure, for a given $M_*$ the radio luminosities of FRII radio 
galaxies with hot spots are systematically higher than those of most of the remaining galaxies. It can also be seen that  the luminosities of sources with hot spots increase with increasing $M_*$. This suggests that the presence of hot spots is not simply
related to a high $P_{\rm 1.4GHz}$. In Fig. \ref{fig:f10} we show the same  sources 
in the $P_{\rm 1.4GHz}$/$``M_{\rm BH}"$ vs $M_*$ diagram. Here, the separation between the two groups of FRII galaxies 
is even more pronounced and one can see that sources with hot spots have larger 
$P_{\rm 1.4GHz}$/$``M_{\rm BH}"$. In other words, hot spot prominence is
connected to higher radio efficiency (and, consequently, a higher Eddington parameter).

\section{Emission line analysis}
\label{sec:emlines}

\begin{figure}
\centering
\includegraphics[width=8cm,angle=0]{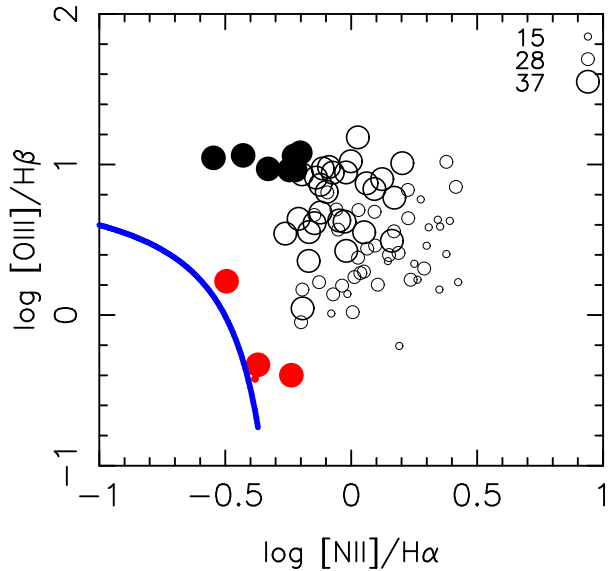}
\caption{FRII radio galaxies in the BPT diagram for objects having S/N $> 3$ in all the relevant lines. The sizes of the symbols correspond to the Eddington parameters  $L_{\Ha}$/$``M_{\rm BH}"$ like in  Fig.   \ref{fig:f7}. A few objects are represented in filled black or grey (red in the electronic version) circles. They are discussed in the text. The thick grey (blue in the electronic version) curve represents the line above which all the galaxies are believed to host an AGN, according to S06.}
\label{fig:f11}
\end{figure}

\begin{figure}
\centering
\includegraphics[width=8cm,angle=0]{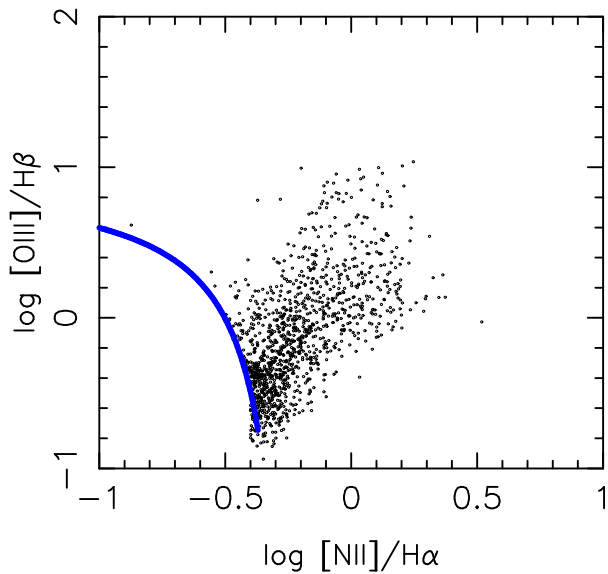}
\caption{The BPT diagram for a comparison sample of 1000 galaxies. These galaxies are randomly drawn from SDSS DR7 galaxies that lie above the S06 line with S/N $> 3$ in all the relevant lines. The thick grey (blue in the electronic version) curve represents the S06 line. }
\label{fig:f12}
\end{figure}

We now investigate in more detail the emission lines in FRII radio galaxies, and in particular the line ratios, in order to get more clues about the characteristics of nuclear activity in these objects. 

\subsection{Diagnostic diagrams}
\label{sec:diagn}

Figure   \ref{fig:f11} plots our FRII radio galaxies in the classical Baldwin, Phillips \& Terlevich (1981, BPT) diagram, used to characterize the ionizing source of the gas in galaxies. We have plotted only the sources with S/N $> 3$ in all the relevant lines, which limits our original sample of FRII galaxies to only 81 objects. The sizes of the circles correspond to the values of 
 $L_{\Ha}$/$``M_{\rm BH}"$ like in  Fig.   \ref{fig:f7}. The galaxies represented with the filled black and grey (red in the electronic version) circles will be discussed in more detail later. The thick grey (blue in the electronic version) curve represents the line above which all the galaxies are believed to host an AGN, according to Stasi\'nska et al. (2006, S06). One can see that all our FRII galaxies lie above the S06 line, as expected. Their distribution in the BPT plane is however very different from that of a random comparison sample of 1000 SDSS DR7 galaxies\footnote{We use comparison samples of limited size rather than the entire SDSS data set to ease visual comparison with our FRII sample in the diagnostic diagrams.} that lie above the S06 line and have S/N $> 3$ in all the relevant lines, as seen in Fig.  \ref{fig:f12}. Most of the objects in the comparison sample gather close to the blue line. In those, star formation is believed to compete with the AGN to produce the emission lines, with the contribution of star formation decreasing away from the blue line (S06). In our FRII sample, there are only a few objects (represented in grey (red in the electronic version)  in Fig.   \ref{fig:f11}) which lie close to the blue line. Most of the FRII galaxies lie well away from it, indicating that the line emission in them is  entirely (or almost entirely) due to their AGN. One can also see that in our FRII sample, there is a much larger proportion of objects having high \oiii/\Hb\ and low \nii/\Ha\ (those represented in black in Fig.  \ref{fig:f11}) than in the comparison sample.
 
\begin{figure}
\centering
\includegraphics[width=8cm,angle=0]{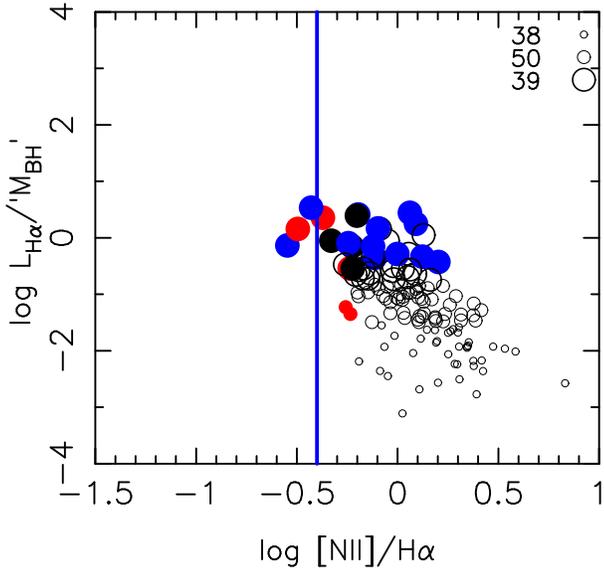}
\caption{FRII radio galaxies in the  $L_{\Ha}$/$``M_{\rm BH}"$ vs \nii/\Ha\ diagram. Only objects with S/N $> 10$ in the continuum and $>3$in \nii\ and \Ha\ are shown. Symbols are like in  Fig.   \ref{fig:f7}, but some of the large open circles have been replaced by filled blue circles (discussed in the text). The  vertical line corresponds to log \nii/\Ha\ $=-0.4$ (see S06).}
\label{fig:f15}
\end{figure}
 
\begin{figure}
\centering
\includegraphics[width=8cm,angle=0]{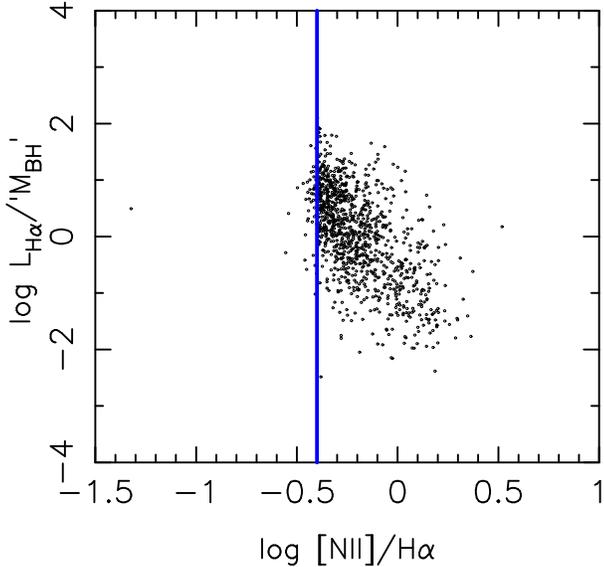}
\caption{The same comparison sample as in \ref{fig:f3} in the  $L_{\Ha}$/$``M_{\rm BH}"$ vs \nii/\Ha\ diagram. Only objects with S/N $> 10$ in the continuum  are shown. The vertical line corresponds to log \nii/\Ha\ $=-0.4$.}
\label{fig:f14}
\end{figure}
 
 Figure   \ref{fig:f15} plots our FRII radio galaxies in the $L_{\Ha}$/$``M_{\rm BH}"$ vs \nii/\Ha\ diagram, which allows one to accomodate 127 of our objects (they must have S/N $> 10$ in the continuum and $>3$ in \nii\ and \Ha), in comparison with the 81 in the BPT diagram. The vertical line has been proposed by Stasi\'nska et al. (2006) to separate galaxies containing an AGN in a more ``economical'' way than the BPT diagram. Some of the grey (red in the electronic version)  and black filled circles actually lie to the left of the vertical line, paradoxically in the region were only ``pure star-forming galaxies'' should lie. We will come back to this below. Figure   \ref{fig:f14} plots our comparison sample in the same diagram, still restricting to galaxies with  S/N $> 10$ in the continuum. Again, the distribution of the comparison sample in this diagram is significantly different from that of our FRII sample. Most of our FRII galaxies, even in this diagram which includes more points thÄan the BPT diagram, still lie far away from the star-forming region. The bulk of the galaxies from the comparison sample, which lie close to the log \nii/\Ha\ $=-0.4$ line, have higher values of $L_{\Ha}$/$``M_{\rm BH}"$. This is because, in the latter objects, $L_{\Ha}$ is affected by photoionization by recently born massive stars.

\subsection{Special cases}
\label{sec:special}
  
 We now turn to discuss the special cases, represented by filled black and grey (red in the electronic version) circles in Figs. \ref{fig:f11} and \ref{fig:f15}.
 
\begin{figure}
\centering
\includegraphics[width=8cm,angle=0]{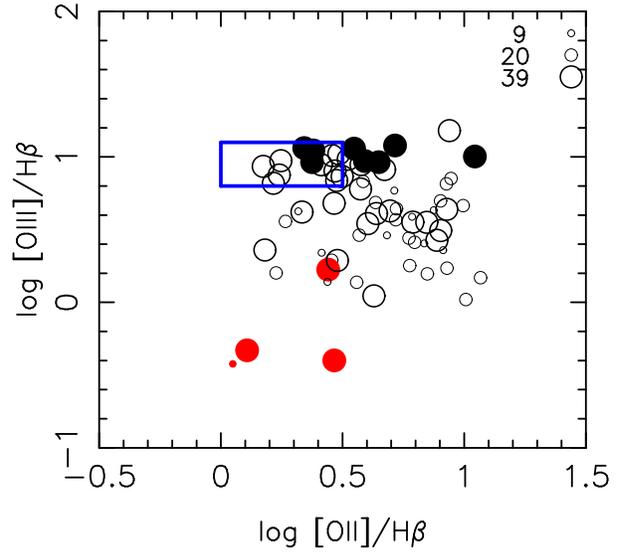}
\caption{FRII radio galaxies in the \oiii/\Hb\ vs \oii/\Hb\  diagram for objects having S/N $> 3$ in all the relevant lines. The sizes of the symbols correspond to the ``Eddington parameters''  $L_{\Ha}$/$``M_{\rm BH}"$ like in  Fig.   \ref{fig:f7}. The objects represented in filled black or grey (red in the electronic version) circles are the same as in Figs.  \ref{fig:f11} and Fig.   \ref{fig:f15}. The  box delimits a zone with log \oiii/\Hb\ $> 0.8$ and log \oii/\Hb\ $> -0.5$, to be compared to Fig. \ref{fig:f17}.}
\label{fig:f16}
\end{figure}

\begin{figure}
\centering
\includegraphics[width=8cm,angle=0]{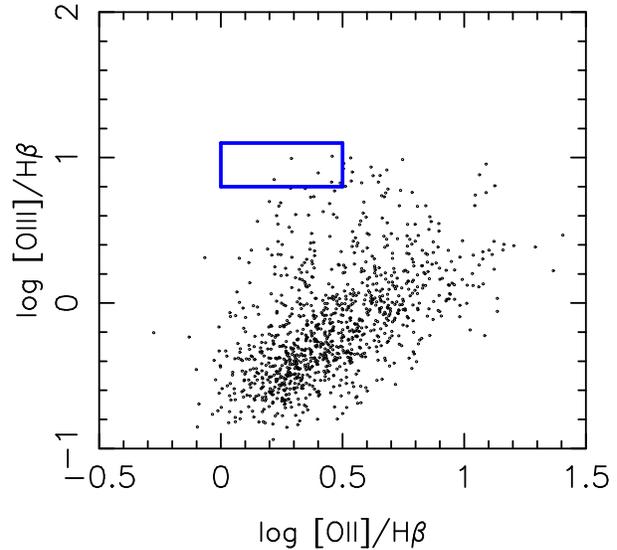}
\caption{The  \oiii/\Hb\ vs \oii/\Hb\  diagram for the comparison sample. The zone delimited by the  box is the same as in Fig. \ref{fig:f16}. }
\label{fig:f17}
\end{figure}

 In the BPT diagram, the position of galaxies containing an AGN is determined by the hardness of the ionizing radiation field, the ionization parameter (which scales like the surface density of ionizing photons per atom of emitting gas) and the O/H and N/O ratios. The distance to the blue curve in Fig. \ref{fig:f11} increases with hardness of the radiation field, while an increase of the ionization parameter moves the points ``parallel'' to the curve towards higher values of \oiii/\Hb. 
 
 The objects represented by black circles in Fig. \ref{fig:f11} are thus likely to be characterized by an exceptionally hard radiation field and a large ionization parameter as compared to the bulk of optical AGN. However it is possible that their position to the left of most galaxies could be due to a small N/O ratio rather than to a large ionization parameter. Figure \ref{fig:f16} shows our FRII radio galaxies in the \oiii/\Hb\ vs \oii/\Hb\  diagram. This diagram is not so efficient as the BPT diagram in distinguishing star forming galaxies from galaxies hosting an AGN, as shown by S06, and therefore is less popular. But if one knows that one is   dealing with AGN galaxies, as is the case here, this diagram is very useful, since it does not depend on the N/O ratio which spans a large range of values in massive galaxies. The points represented in black in Fig. \ref{fig:f16} are the same as those represented in black in Figs.  \ref{fig:f11} and  \ref{fig:f15}. One can therefore infer that many of them have actually a small N/O ratio. On the other hand, the objects that, schematically,  are found in the  box in Fig. \ref{fig:f16} certainly have both a very hard ionizing radiation field and a high ionization parameter. There are 13 such objects out of 69 objects appearing in  Fig. \ref{fig:f16}, while there are only 8 objects out of nearly 1000 in the  box in  Fig. \ref{fig:f17} constructed with the comparison sample. Thus, the sample of FRII radio galaxies is characterized by a large proportion of AGNs with the hardest ionizing radiation field and highest ionization parameters. The objects that appear in the blue box in Fig. \ref{fig:f16} are represented with filled grey (blue in the electronic version)  circles in Fig.  \ref{fig:f15} (unless they were already black).  Fig.  \ref{fig:f15} thus shows that the objects with the highest ionization parameters also belong to the ones with the largers values of the Eddington parameter,  $L_{\Ha}$/$``M_{\rm BH}"$. 
 
\begin{figure}
\centering
\includegraphics[width=8cm,angle=0]{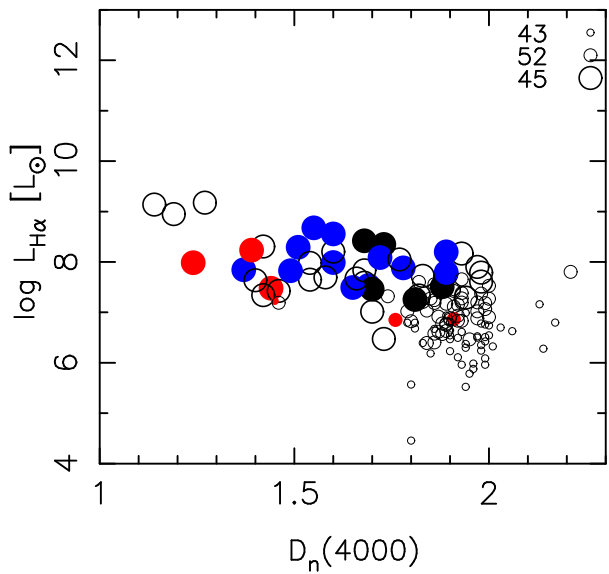}
\caption{FRII radio galaxies in the  $L_{\Ha}$ vs $D_{\rm n}(4000)$ plane. Only objects with S/N $> 10$ in the continuum are shown. The symbols are exactly the same as in  Fig.   \ref{fig:f11}.}
\label{fig:f18}
\end{figure}

\begin{figure}
\centering
\includegraphics[width=8cm,angle=0]{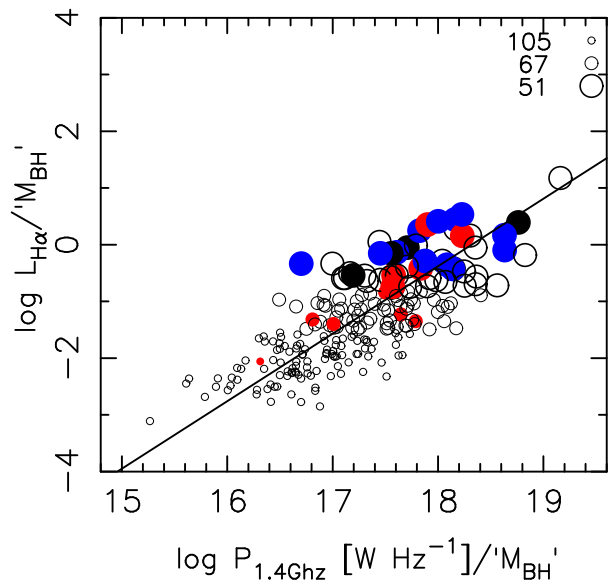}
\caption{FRII radio galaxies in the  $L_{\Ha}$/$``M_{\rm BH}"$ vs $P_{\rm 1.4GHz}/``M_{\rm BH}"$ plane. Only objects with S/N $> 10$ in the continuum are shown. The symbols are exactly the same as in  Fig.   \ref{fig:f15}.}
\label{fig:f19}
\end{figure}

\begin{figure}
\centering
\includegraphics[width=5cm,angle=0]{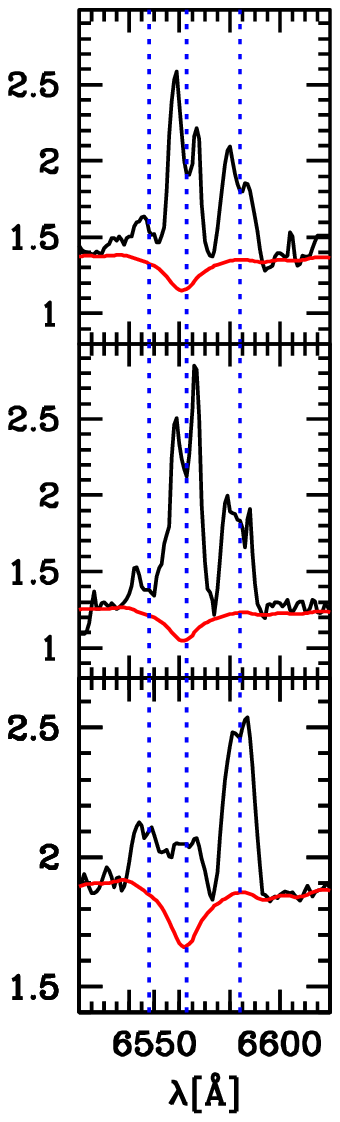}
\caption{Spectra of the objects with double lines in the \Ha\ and \nii\ region. Top: object 0818.52395.570, one of the grey (red in the electronic version) points in Fig. \ref{fig:f11}.
Middle: object 0902.52409.325, one of the grey (red in the electronic version)  points in Fig. \ref{fig:f11}. Bottom: object 0932.52620.071, the third of our FRII galaxies with possible double lines.}
\label{fig:f20}
\end{figure}

 The nature of the objects represented by grey (red in the electronic version)  circles in Figs.  \ref{fig:f11}, \ref{fig:f15} and  \ref{fig:f16} is more difficult to understand. Those objects are close to the blue curve in the BPT diagram, so one could think they belong to the class of so-called composite galaxies, where the ionization by young massive stars competes with that of the AGN. If this is the case, one would expect the break at 4000\AA, $D_{\rm n}(4000)$, to have a low value, of the order of 1--1.5, indicating the presence of young stellar populations (Bruzual 1983; Cid Fernandes et al. 2005).   Figure \ref{fig:f18} shows the positions of those galaxies with respect to the other ones from our FRII sample in the $L_{\Ha}$ vs $D_{\rm n}(4000)$ plane.  Here, we use the definition of Balogh et al. (1999) for  $D_{\rm n}(4000)$ and consider only spectra with a S/N $>10$ in the continuum. It can be seen that, indeed, those objects are among the ones with the youngest stellar populations. 
 
 Since, in the BPT diagram, and also in the \oiii/\Hb\ vs \oii/\Hb\ diagram, the objects lie so much to the left of the remaining ones, the contribution of massive star ionization should be by far dominant. Following the composite photoionization models by S06 the contribution of the AGN to \Ha\ should be of only 3\%. In such a case, those objects should stand out from the relation between  $L_{\Ha}$/$``M_{\rm BH}"$ and $P_{\rm 1.4GHz}/``M_{\rm BH}"$. We show again the  $L_{\Ha}$/$``M_{\rm BH}"$ vs $P_{\rm 1.4GHz}/``M_{\rm BH}"$ plot for our FRII radio galaxies, this time using the same symbols as in Fig. \ref{fig:f15}. We see that, although the grey (red in the electronic version)  points are found above the mean of the $L_{\Ha}$/$``M_{\rm BH}"$ vs $P_{\rm 1.4GHz}/``M_{\rm BH}"$ relation, they do not stand out particularly with respect to the remaining objects. Therefore, we are inclined to think that the location of the grey (red in the electronic version)  points in the emission-line diagnostic diagrams, rather than being due to an overwhelming presence of young stars, should be attributed to a much softer ionizing radiation  as compared with the rest of the AGNs.
 
 There is another intriguing characteristc of those objects. Two of them, 0818.52395.570 and 0902.52409.325 seem to have double line profiles, as can be seen in Fig. \ref{fig:f20}. There is only one other object with such profiles in our sample of FRII galaxies, also shown in Fig. \ref{fig:f20}.  According to Liu et al. (2010a and b), such features are likely due to massive  binary black holes. Clearly, those objects deserve further, more detailed observations, to uncover their real nature. 
 

\section{FRII galaxies in perspective}
\label{sec:loud-quiet}

\begin{figure}
\centering
\includegraphics[width=8cm,angle=0]{FRII_ufrii.ps}
\caption{The rest-frame (u-r) color as a function of $M_{*}$ for the FRII galaxies with $z<0.1$. The symbol sizes represent correspond to the values of log $L_{\Ha}$/$``M_{\rm BH}"$, as in  Fig. \ref{fig:f6}.}
\label{fig:ufrii}
\end{figure}

\begin{figure}
\centering
\includegraphics[width=8cm,angle=0]{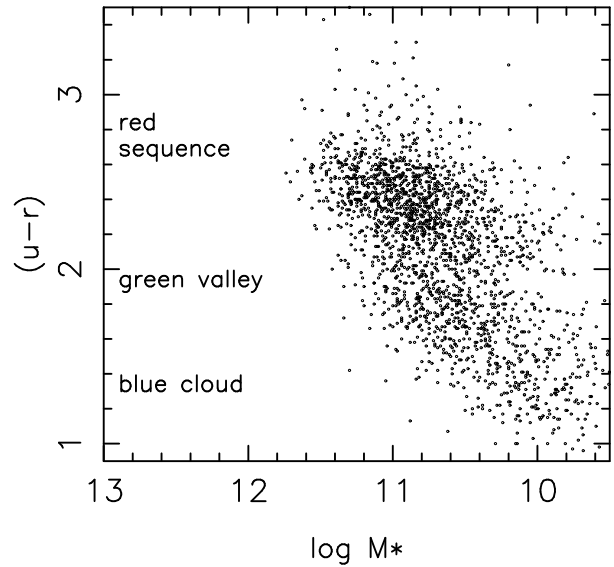}
\caption{The rest-frame (u-r) color as a function of $M_{*}$ for a random comparison sample of 2000 objects in the SDSS with  with $z<0.1$.}
\label{fig:uall}
\end{figure}

\begin{figure*}
\centering
\includegraphics[width=17cm,angle=0]{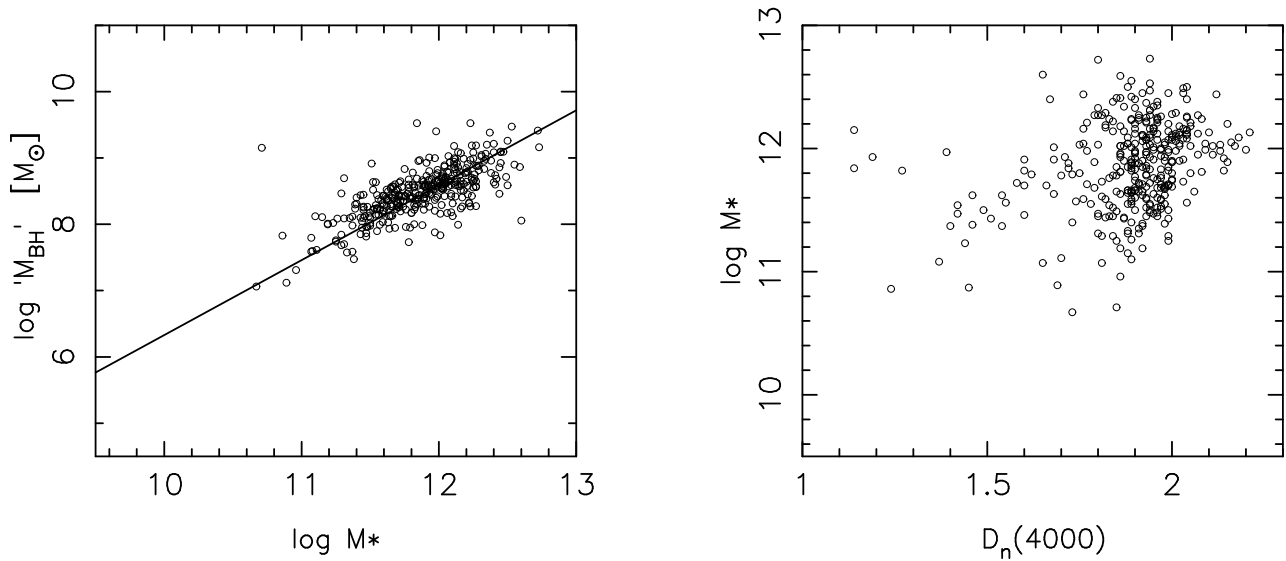}
\caption{$``M_{\rm BH}"$ vs $M_{*}$ (left) and $M_{*}$  vs  $D_{\rm n}(4000)$ (right) for our sample of FRII galaxies. The straight line is the bissector regression line given by Eq. \ref{eq:MM}. }
\label{fig:frii}
\end{figure*}

\begin{figure*}
\centering
\includegraphics[width=17cm,angle=0]{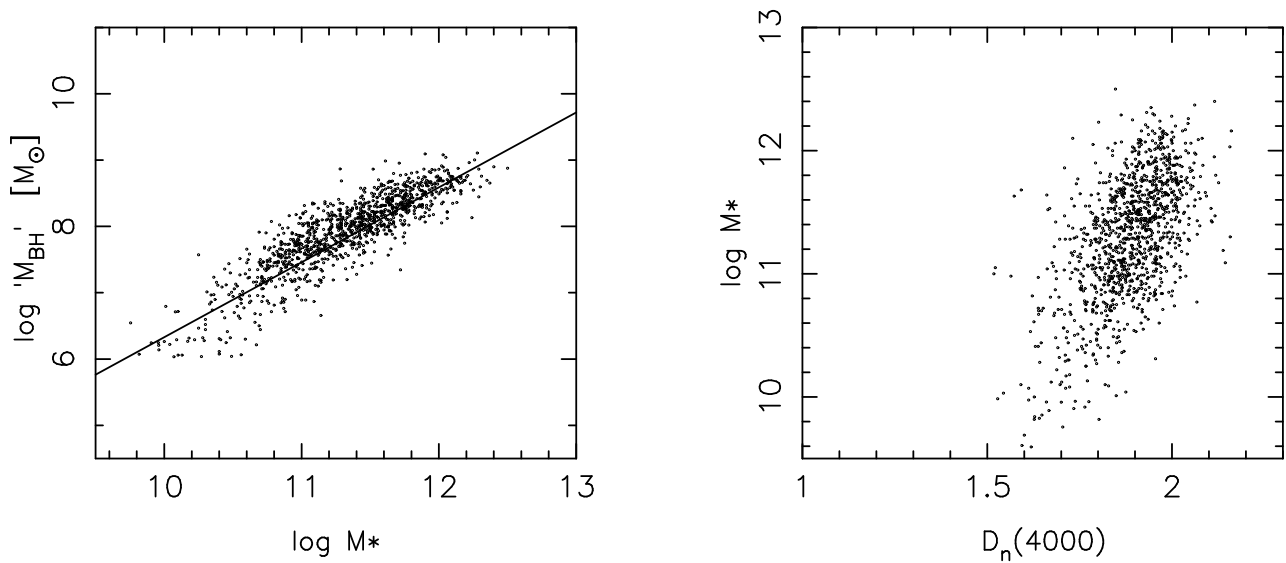}
\caption{The same as  Fig. \ref{fig:frii} but  for a random sample  of 1000 line-less galaxies in the SDSS. The straight line is the same as in Fig. \ref{fig:frii}.}
\label{fig:lineless}
\end{figure*}

\begin{figure*}
\centering
\includegraphics[width=17cm,angle=0]{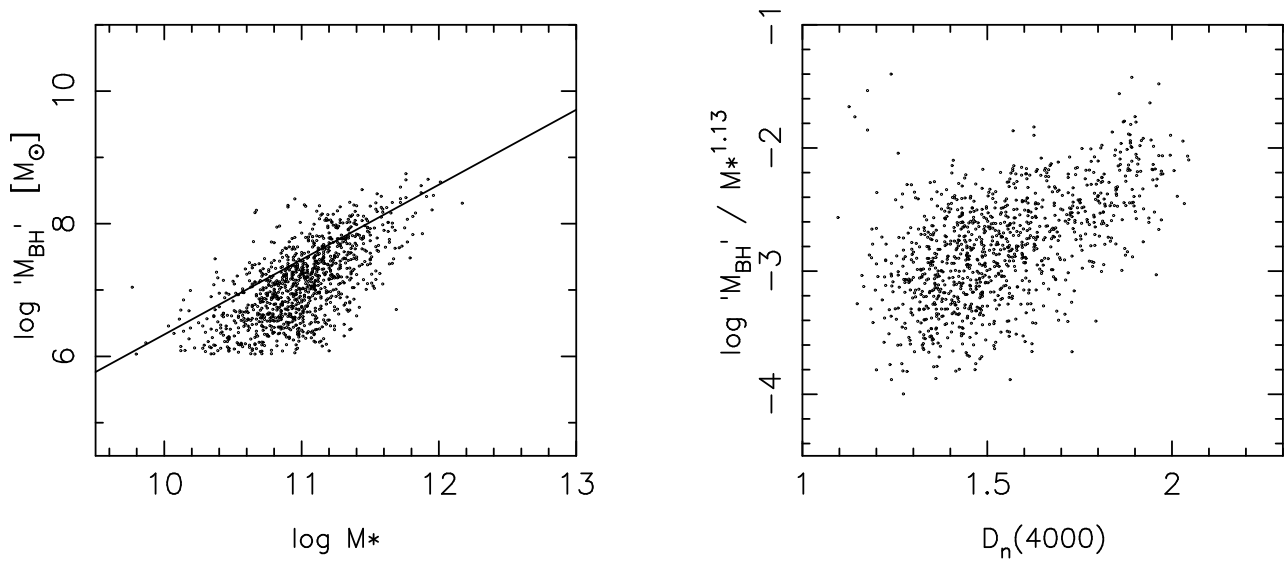}
\caption{$``M_{\rm BH}"$ vs $M_{*}$ (left) and   $``M_{\rm BH}"$/$M_{*}^{1.13}$  vs  $D_{\rm n}(4000)$ (right) for a random sample of 1000 galaxies considered to contain an AGN (i.e. galaxies that lie above the S06 curve in the BPT diagram). The straight line is the same as in Fig. \ref{fig:frii}.}
\label{fig:agn}
\end{figure*}

We now characterize our population of FRII radio galaxies with respect to 
populations of related objects analyzed with the same procedures.

 In Fig. \ref{fig:ufrii} we show the position of our FRII galaxies in the classical color-mass, (u-r) vs $M_{*}$, diagram. Only the 57 objects with redhifts $z<1$ are shown, for a sound comparison with Fig. 4 of Smol{\v c}i{\'c} (2009).  This, of course,  suppresses  the most luminous and massive of our FRII radio galaxies, as can be seen when comparing Fig. \ref{fig:ufrii} with Fig.  \ref{fig:f6}. Still, the objects from our FRII sample that remain in  Fig. \ref{fig:ufrii} tend to have higher masses than those appearing in Fig. 4 of  Smol{\v c}i{\'c} (2009), due to the fact that the latter is based on a sample likely dominated by  galaxies associated to compact or FRI radio sources. For comparison, in Fig. \ref{fig:uall} we plot  a random sample of 2000 galaxies of all types from the SDSS (again with redshifts  $z<1$). In both Figs. \ref{fig:ufrii} and \ref{fig:uall}, we indicate the approximate colour of the red sequence, green valley and blue cloud, following the nomenclature of Bell et al. (2004). In Fig. \ref{fig:ufrii} the symbol sizes represent the values of   $L_{\Ha}$/$``M_{\rm BH}"$, like in  Fig. \ref{fig:f6}. We can see that the FRII  galaxies with lowest Eddington parameters fall in the region of the red sequence, while those with the largest Eddington parameters fall in the region of the green valley. This reminds the result of Smol{\v c}i{\'c} (2009) for low- and high-excitation AGNs of various radio morphologies. This does not mean, however, that  high-excitation (or high $L_{\Ha}$/$``M_{\rm BH}"$) radio galaxies are not intrinsically red. As a matter of fact, the spectra of most of our FRII galaxies, including most of those with high Eddington parameters,  have features similar to those of red galaxies.  Simply, the effect of the active nucleus is to bias the photometric colour of the host galaxy, as shown for example by  Pierce et al. (2010). 
  
In Fig. \ref{fig:frii} we represent our FRII sample in the   $``M_{\rm BH}"$ vs $M_{*}$ plot (left) and in the $M_{*}$  vs  $D_{\rm n}(4000)$ plot (right). 

 Figure \ref{fig:lineless} shows the same, but for a random sample of 1000 line-less galaxies in the SDSS (S/N $<2$ in all major emission lines). Those galaxies are typical red  galaxies that show no sign of activity. 
 
  The correlation between $``M_{\rm BH}"$ and $M_{*}$ for FRII galaxies is very strong ($R$=0.70). The bissector regression line, assuming errors of 0.2 dex on each quantity, is given by:
 \begin{eqnarray}
{\rm log} ``M_{\rm BH}" = (1.13 \pm 0.058)  \times {\rm log} M_{*} -   \nonumber\\
(4.97 \pm 0.69).
\label{eq:MM}
\end{eqnarray} 

It is plotted  in Fig. \ref{fig:frii}, as well as in Fig.  \ref{fig:lineless}, for reference. Clearly, this line is also a good fit to the relation observed between $``M_{\rm BH}"$ and stellar mass in line-less galaxies. Here, the correlation coefficient is even as large as  $R$=0.85. 
We may note that the slope of the   log $``M_{\rm BH}"$ -- log $M_{*}$ relation is slightly, 
but significantly, smaller than one. 
One can notice, also, that the distribution of stellar masses and ``black hole masses'' are not 
exactly the same in the two samples: they are shifted towards higher values in the FRII sample. 
This implies that the FRII phenomenon does not occur in less massive line-less galaxies.

In the $M_{*}$  vs  $D_{\rm n}(4000)$ diagram, although most  FRII galaxies are old, as indicated by the $D_{\rm n}(4000)$ index already commented before, there are a few galaxies which have  $D_{\rm n}(4000)$ $<$ 1.5, indicating the presence of young stellar populations (with ages of the order of $10^7 - 10^8$~yr). There is no hint of such populations in the comparison sample of line-less galaxies, as seen in Fig.  \ref{fig:lineless}. Note that, in the FRII sample, young stellar populations can occur at any value of $M_{*}$. This suggests that there is a connection between this recent star formation, presumably due to a collision, and the optical and radio activity.

If we now compare  Fig.  \ref{fig:frii} (left) with Fig.  \ref{fig:agn} (left), which is identical 
to Fig.  \ref{fig:lineless} (left) but for our comparison sample of galaxies that lie above the blue curve in the BPT diagram (Fig. \ref{fig:f11}), we see that $``M_{\rm BH}"$ and  $M_{*}$ are not so well correlated as in the previous sample\footnote{The lower envelope of the points in Fig. \ref{fig:agn} (left) is artificial and corresponds to the minimum value of $\sigma_*$ suitable to compute $``M_{\rm BH}"$.}. In this case, we are mostly dealing with spiral galaxies, where the bulge is not very prominent, so the estimate of the black hole mass using the measured stellar velocity dispersion might be biased.  However,  when excluding galaxies with redshift $> 0.1$, so that the measurement of $\sigma_{*}$ is not strongly affected by the galaxy disk, we obtain the same picture. The bulk of objects deviate from the  $``M_{\rm BH}"$ -- $M_{*}$ relation obtained for FRII or line-less galaxies. For a given galaxy mass, $``M_{\rm BH}"$ tends to be smaller. This suggests that those galaxies, which are still forming stars, are also still  building their central massive black hole. To test this suggestion, we plot, in Fig.  \ref{fig:agn} (right), the values of  $``M_{\rm BH}"$/$M_{*}^{1.13}$ (which is constant in the more evolved, line-less galaxies of Fig. \ref{fig:lineless} as a function of $D_{\rm n}(4000)$ for the same sample as in in Fig.  \ref{fig:agn} (left). We clearly see a that younger, i.e. less evolved galaxies that have smaller $D_{\rm n}(4000)$ tend to have smaller   $``M_{\rm BH}"$/$M_{*}^{1.13}$ , and that this ratio increases with  $D_{\rm n}(4000)$.

\section{Summary}

Using the Cambridge Catalogues of radio sources we have built a sample of FRII radio galaxies whose spectra are available in the main galaxy sample of SDSS DR7. From this sample, after inspection of the NVSS and FIRST radio maps, we extracted a sample of 401 FRII radio galaxies. In this paper, we examined the optical and radio properties of those objects, in order to find new clues about the relation between radio activity and the optical manifestations of the AGN in those objects. 
The  stellar masses,  emission line equivalent widths and fluxes were taken from the \starlight\ database
(Cid Fernandes et al. 2009). 

We found that the luminosity in the H$\alpha$ line -- which we argue gives a better measure of the total flux in the emission lines than the widely used luminosity in \oiii\ -- is strongly correlated with the radio luminosity $P_{\rm 1.4GHz}$ over more than three orders of magnitude in  $P_{\rm 1.4GHz}$. A similar result was  found by Zirbel \& Baum (1995), using \Ha\ + \nii, but they obtained  $L_{\Ha+\nii}$ $\propto$ $P_{\rm 1.4GHz}^{0.75}$ while we obtain $L_{\Ha}$ $\propto$ $P_{\rm 1.4GHz}^{1.13}$.  In our sample, there is about one third of objects which do not have any line detected. We showed that, for those,  the detection threshold is above the empirical relations between $L_{\oiii}$ or $L_{\Ha}$ and $P_{\rm 1.4GHz}$. Therefore, there is nothing for the moment that indicates the existence of two classes of FRII radio galaxies with respect to the presence  of emission lines. 

We also find a very strong correlation between the values of $L_{\Ha}$ and $P_{\rm 1.4GHz}$ when scaled by $``M_{\rm BH}"$, the black hole masses obtained from the observed stellar velocity dispersion suggesting that, in FRII radio galaxies, optical and radio activity have a common cause.

 Contrary to previous work (e.g. Buttiglione et al 2010, or Lin et al. 2010, however based on different samples) we see no sharp transition  between high- and low-excitation radio galaxies or galaxies with $L_{\oiii}$ smaller or larger than $10^6 L_\odot$. We rather find that FRII galaxies present a continuum of properties driven by the Eddington parameter  $L_{\Ha}$/$``M_{\rm BH}"$ or, equivalently,  $P_{\rm 1.4GHz}$/$``M_{\rm BH}"$, 
 We note, however, that FRII galaxies with hot spots are found among the ones with the highest values of $P_{\rm 1.4GHz}$/$``M_{\rm BH}"$. Those hot spots, which are the places where the relativistic jets from the active galactic nuclei interact with the environment, thus seem to require a radiatively efficient accretion to be produced.

Those FRII galaxies that can be plotted in the classical BPT diagram or similar diagrams fall in the zone characterized by a hard ionizing spectrum, where the contribution of hot stars to the excitation is small or absent. This is expected, since it is known  that radio galaxies are in general associated with massive, elliptical galaxies. Compared to classical AGN hosts found in the galaxy sample of the SDSS, there is a significantly larger proportion of objects with very hard ionizing radiation field and large ionization parameter. There are, however a few objects that lie close to the divisory line between pure star-forming galaxies and AGN hosts. This is a priori surprising, as there is no indication of present-day star formation in those objects. Two of them have  double-peaked lines, as are found in some  AGNs and attributed to  binary black holes (among other possibilities). We suggest that those objects are ionized by a rather soft radiation field, as compared with the rest of the FRIIs.

For the 346 FRII galaxies for which we could determine both the black hole mass and the stellar mass, we find that $``M_{\rm BH}"$ varies like $M_{*}^{1.13}$ with very little scatter. A comparison sample of line-less galaxies in the SDSS follows exactly the same relation, but with both masses shifted to lower values. This suggests that the FRII radio phenomenon occurs in normal elliptical galaxies, but is favoured by larger galaxy masses. The $D_{\rm n}(4000)$ index indicates that, although most of the FRII galaxies are old, some  contain traces of young stellar populations. Since such young populations are not seen in normal line-less galaxies, one can conjecture that the radio (and optical) activity is triggered by recent star formation. The $``M_{\rm BH}"$ -- $M_{*}$ relation in a comparison sample of radio-quiet AGN hosts from the SDSS is very different, suggesting that galaxies which are presently forming stars are still building their central black hole. The  $D_{\rm n}(4000)$ index in this sample indicates that the youngest galaxies have the smallers  $``M_{\rm BH}"$/$M_{*}$ ratio, confirming this view.

Overall, our study leads to the conclusion that, while radio and optical activity are strongly related in FRII galaxies, the features of the optical activity in those objects are distinct from those of radio quiet active galaxies.

\section*{ACKNOWLEDGMENTS}

 This work was carried out within the framework of the European Associated Laboratory "Astrophysics Poland-France". LEA Astro-PF. 
 DKW was partialy supported by the MNiSW with funding for the scientific research in years 2009-2012 under contract No. 3812/B/H03/2009/36.
 We thank R. Cid Fernandes and associates, especially Natalia Vale Asari and William Schoenell, 
 to have made available the \starlight\ data base and helped us to make good use of it. We thank Marek Sikora and the referee for good suggestions.
DKW thanks prof. Jerzy Machalski for arising her interest in radio galaxies. G.S. thanks S. Collin-Zahn for stimulating discussions.
The Sloan Digital Sky Survey is a joint project of The University of
Chicago, Fermilab, the Institute for Advanced Study, the Japan
Participation Group, the Johns Hopkins University, the Los Alamos
National Laboratory, the Max-Planck-Institute for Astronomy, the
Max-Planck-Institute for Astrophysics, New Mexico State University,
Princeton University, the United States Naval Observatory, and the
University of Washington.  Funding for the project has been provided
by the Alfred P. Sloan Foundation, the Participating Institutions, the
National Aeronautics and Space Administration, the National Science
Foundation, the U.S. Department of Energy, the Japanese
Monbukagakusho, and the Max Planck Society.

\appendix
\onecolumn

\section[]{Catalogue of the Cambridge-SDSS FRII radio galaxies}

\scriptsize
\begin{center}
\begin{longtable}[t]{r c c c c c c c c c}
\caption{Radio properties. Galaxies with double-peak emission lines are marked with *.} \\
\hline \hline \\[-2ex]
\multicolumn{1}{r}{No.} &
\multicolumn{1}{c}{SDSS ID} & 
\multicolumn{1}{c}{Name} & 
\multicolumn{1}{c}{redshift} & 
\multicolumn{1}{c}{log $S_{1.4GHz}$} & 
\multicolumn{1}{c}{log $L_{1.4GHz}$} & 
\multicolumn{1}{c}{angular size} & 
\multicolumn{1}{c}{linear size} & 
\multicolumn{1}{c}{core} & 
\multicolumn{1}{c}{hot spots} \\
\multicolumn{1}{c}{} & \multicolumn{1}{c}{Plate.MJD.Fiber}  & \multicolumn{1}{c}{} & \multicolumn{1}{c}{} & \multicolumn{1}{c}{[mJy]} & \multicolumn{1}{c}{[W Hz$^{-1}$]} & \multicolumn{1}{c}{[arcsec]} & \multicolumn{1}{c}{[kpc]} & \multicolumn{1}{c}{} & \multicolumn{1}{c}{} \\
\hline
\endfirsthead

\multicolumn{9}{c}{{\tablename} \thetable{} -- continued}\\
\hline \hline
\multicolumn{1}{r}{No.} &
\multicolumn{1}{c}{SDSS ID} & 
\multicolumn{1}{c}{Name} & 
\multicolumn{1}{c}{redshift} & 
\multicolumn{1}{c}{log $S_{1.4GHz}$} & 
\multicolumn{1}{c}{log $L_{1.4GHz}$} & 
\multicolumn{1}{c}{angular size} & 
\multicolumn{1}{c}{linear size} & 
\multicolumn{1}{c}{core} & 
\multicolumn{1}{c}{hot spots} \\
\multicolumn{1}{c}{} & \multicolumn{1}{c}{Plate.MJD.Fiber}  & \multicolumn{1}{c}{} & \multicolumn{1}{c}{} & \multicolumn{1}{c}{[mJy]} & \multicolumn{1}{c}{[W Hz$^{-1}$]} & \multicolumn{1}{c}{[arcsec]} & \multicolumn{1}{c}{[kpc]} & \multicolumn{1}{c}{} & \multicolumn{1}{c}{} \\
\hline
\endhead

\hline \\
\endfoot

                  1&                  0273.51957.633     &                  J103605+000606     &                                  0.1&                                544.6&                                25.65&                                  144&                               258.07&                  y                  &                  -                  \\                 
                  2&                  0281.51614.562     &                  J113021+005823     &                                 0.13&                                666.6&                                26.01&                                   47&                                110.6&                  -                  &                  y                  \\                 
                  3&                  0349.51699.169     &                  J165847+625624     &                                 0.11&                                  286&                                25.45&                                139.2&                                270.9&                  -                  &                  y                  \\                 
                  4&                  0358.51818.161     &                  J173250+563426     &                                 0.33&                                 45.3&                                25.66&                                   30&                               143.41&                  y                  &                  y                  \\                 
                  5&                  0359.51821.502     &                  J172749+534651     &                                 0.21&                               1040.5&                                 26.6&                                146.4&                               496.37&                  y                  &                  y                  \\
\hline                                                                         
\end{longtable}
\end{center}

\section[]{Atlas of the Cambridge-SDSS FRII radio galaxies}

\begin{figure*}
\centering
\includegraphics[width=15cm]{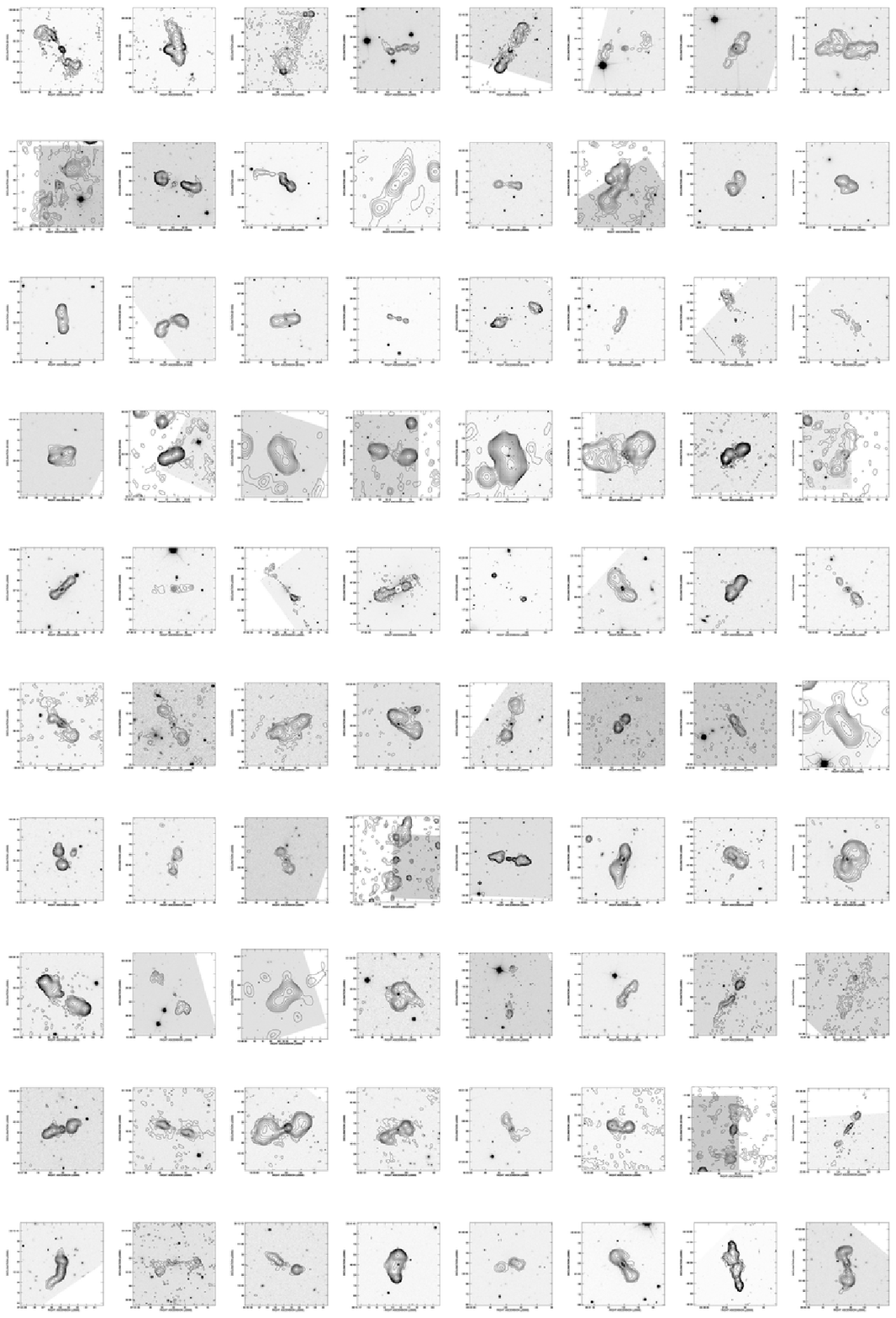}
\caption{FIRST contour radio maps overplotted on gray scale SDSS r-band images of the first 80 objects from the sample. } 
\end{figure*}

\begin{figure*}
\centering
\includegraphics[width=15cm]{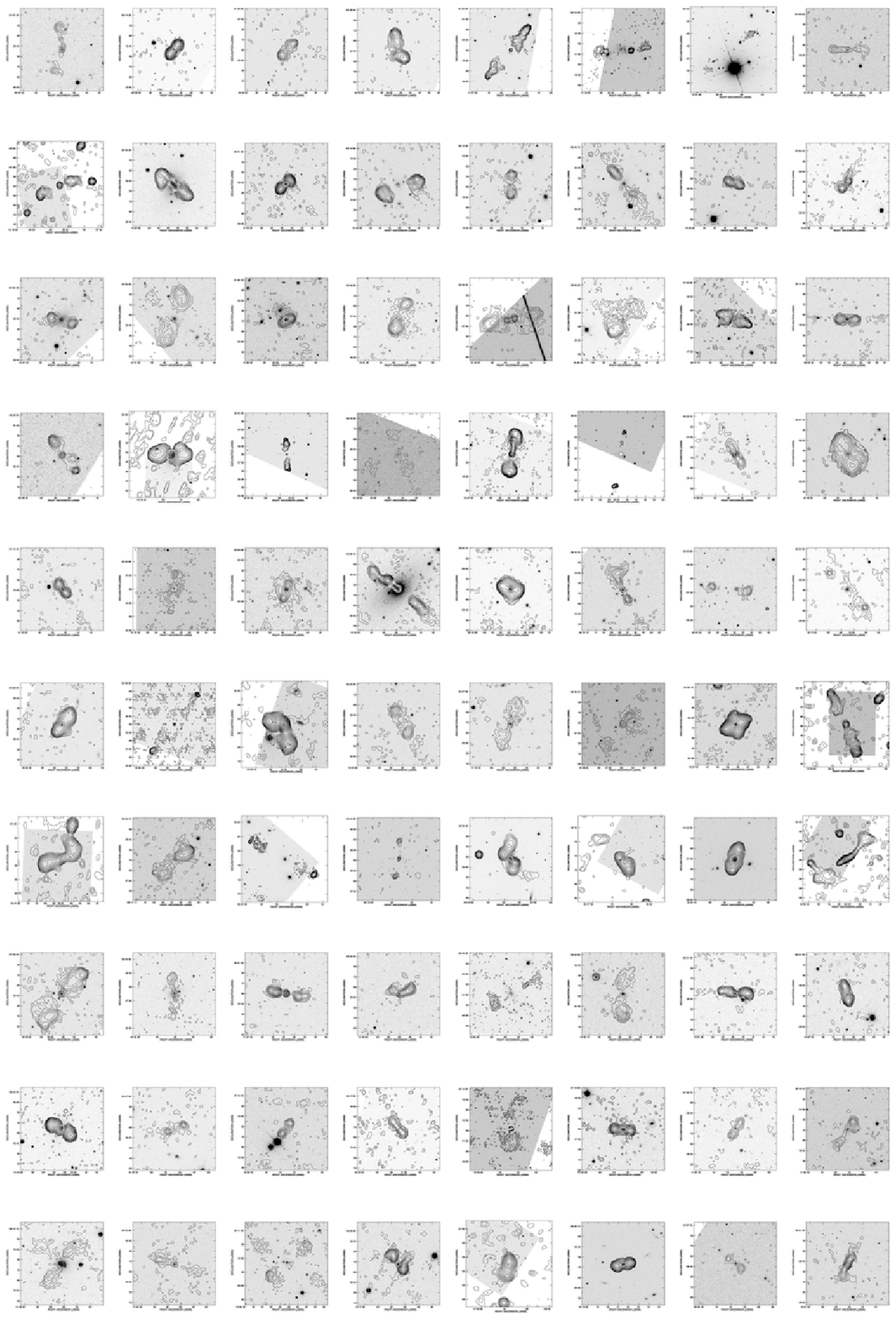}
\caption{FIRST contour radio maps overploted on gray scale SDSS r-band images of next 80 objects from the sample.}
\end{figure*}

\begin{figure*}
\centering
\includegraphics[width=15cm]{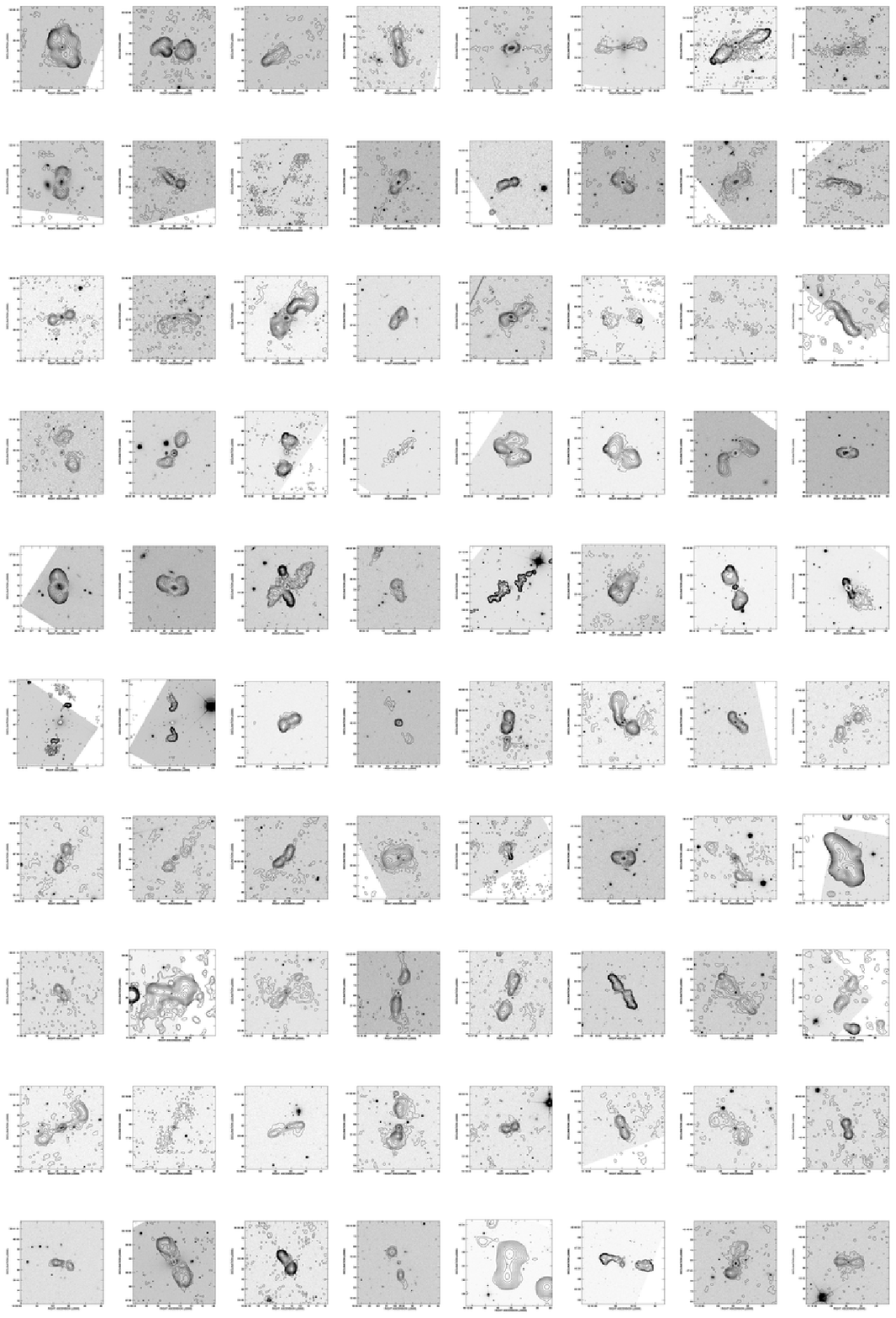}
\caption{FIRST contour radio maps overploted on gray scale SDSS r-band images of the next 80 objects from the sample.}
\end{figure*}

\begin{figure*}
\centering
\includegraphics[width=15cm]{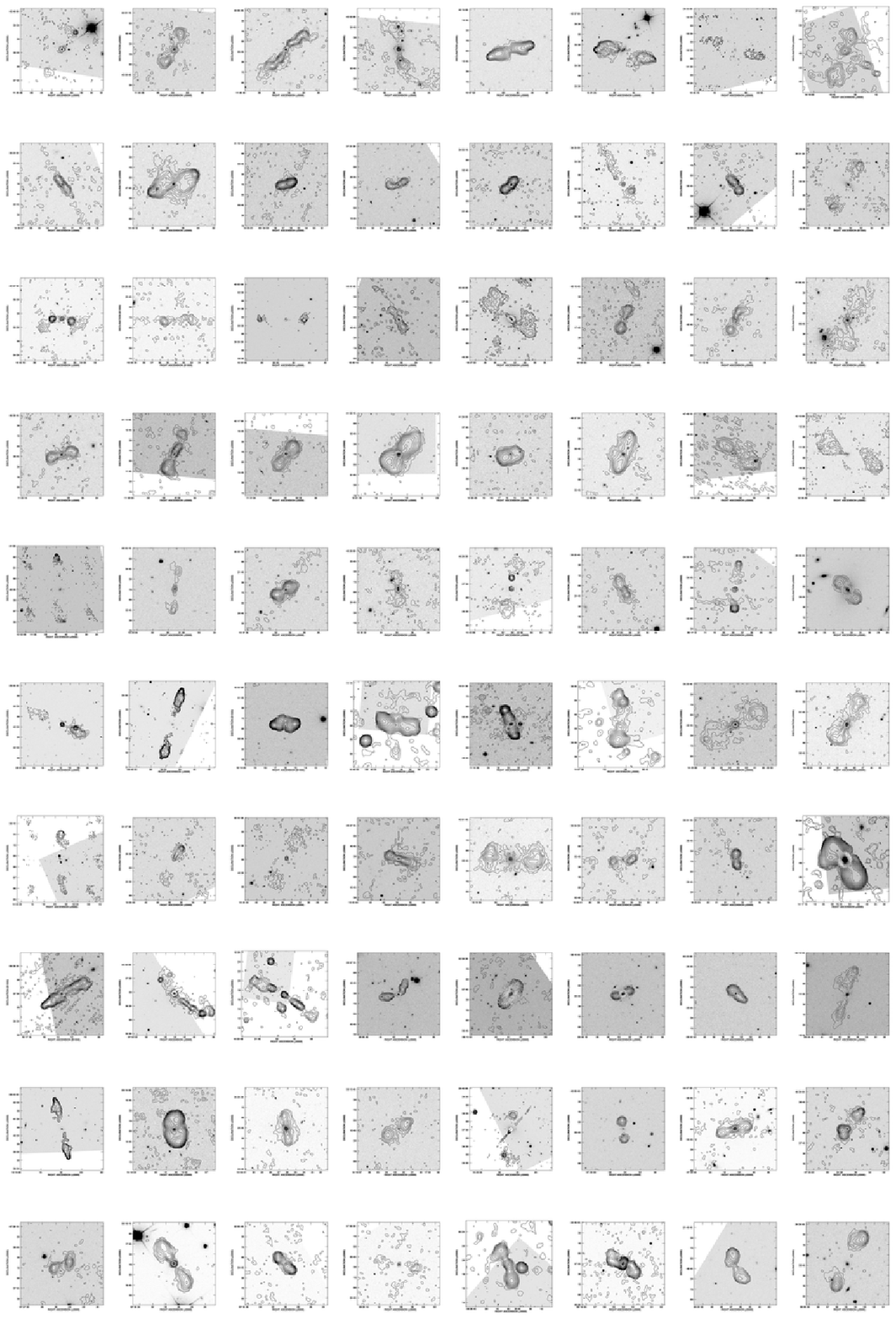}
\caption{FIRST contour radio maps overploted on gray scale SDSS r-band images of the next 80 objects from the sample.}
\end{figure*}

\begin{figure*}
\centering
\includegraphics[width=15cm]{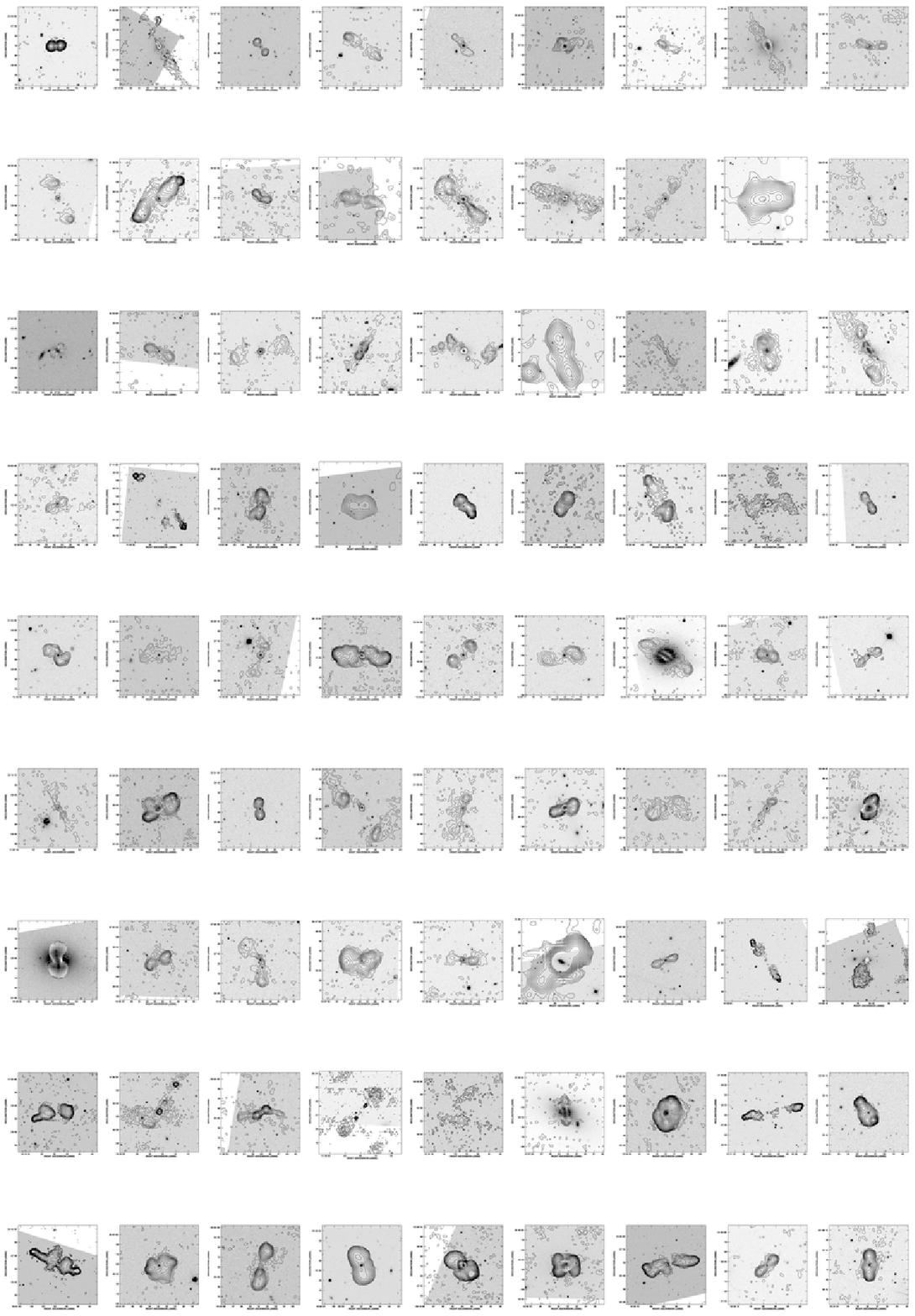}
\caption{FIRST contour radio maps overploted on gray scale SDSS r-band images of the last 81 objects from the sample.}
\end{figure*}


\begin{thebibliography}{}

\bibitem[\protect\citeauthoryear{Abazajian et 
al.}{2004}]{2004AJ....128..502A} Abazajian K., et al., 2004, AJ, 128, 502 

\bibitem[\protect\citeauthoryear{Abazajian et 
al.}{2009}]{2009ApJS..182..543A} Abazajian K.~N., et al., 2009, ApJS, 182, 
543 

\bibitem[\protect\citeauthoryear{Akritas 
\& Bershady}{1996}]{1996ApJ...470..706A} Akritas M.~G., Bershady M.~A., 1996, ApJ, 470, 706 

\bibitem[\protect\citeauthoryear{Baldwin et 
al.}{1985}]{1985MNRAS.217..717B} Baldwin J.~E., Boysen R.~C., Hales 
S.~E.~G., Jennings J.~E., Waggett P.~C., Warner P.~J., Wilson D.~M.~A., 
1985, MNRAS, 217, 717 

\bibitem[\protect\citeauthoryear{Baldwin, Phillips, 
\& Terlevich}{1981}]{1981PASP...93....5B} Baldwin J.~A., Phillips M.~M., Terlevich R., 1981, PASP, 93, 5 

\bibitem[\protect\citeauthoryear{Balogh et al.}{1999}]{1999ApJ...527...54B} 
Balogh M.~L., Morris S.~L., Yee H.~K.~C., Carlberg R.~G., Ellingson E., 
1999, ApJ, 527, 54 

\bibitem[\protect\citeauthoryear{Becker, White, 
\& Helfand}{1995}]{1995ApJ...450..559B} Becker R.~H., White R.~L., Helfand D.~J., 1995, ApJ, 450, 559 

\bibitem[\protect\citeauthoryear{Bell et al.}{2004}]{2004ApJ...608..752B} 
Bell E.~F., et al., 2004, ApJ, 608, 752 

\bibitem[\protect\citeauthoryear{Benn}{1995}]{1995MNRAS.272..699B} Benn 
C.~R., 1995, MNRAS, 272, 699 

\bibitem[\protect\citeauthoryear{Benn 
\& Kenderdine}{1991}]{1991MNRAS.251..253B} Benn C.~R., Kenderdine S., 1991, MNRAS, 251, 253 

\bibitem[\protect\citeauthoryear{Benn et al.}{1982}]{1982MNRAS.200..747B} 
Benn C.~R., Grueff G., Vigotti M., Wall J.~V., 1982, MNRAS, 200, 747 

\bibitem[\protect\citeauthoryear{Bennett}{1962}]{1962MmRAS..68..163B} 
Bennett A.~S., 1962, MmRAS, 68, 163 

\bibitem[\protect\citeauthoryear{Bertelli et 
al.}{1994}]{1994A&AS..106..275B} Bertelli G., Bressan A., Chiosi C., Fagotto F., Nasi E., 1994, A\&AS, 106, 275 


\bibitem[\protect\citeauthoryear{Best et al.}{2005a}]{2005MNRAS.362....9B} 
Best P.~N., Kauffmann G., Heckman T.~M., Ivezi{\'c} {\v Z}., 2005, MNRAS, 
362, 9 

\bibitem[\protect\citeauthoryear{Best et al.}{2005b}]{2005MNRAS.362...25B} 
Best P.~N., Kauffmann G., Heckman T.~M., Brinchmann J., Charlot S., 
Ivezi{\'c} {\v Z}., White S.~D.~M., 2005, MNRAS, 362, 25 

\bibitem[\protect\citeauthoryear{Bruzual A.}{1983}]{1983ApJ...273..105B} 
Bruzual A.~G., 1983, ApJ, 273, 105 

\bibitem[\protect\citeauthoryear{Bruzual 
\& Charlot}{2003}]{2003MNRAS.344.1000B} Bruzual G., Charlot S., 2003, MNRAS, 344, 1000 

\bibitem[\protect\citeauthoryear{Buttiglione et 
al.}{2010}]{2010A&A...509A...6B} Buttiglione S., Capetti A., Celotti A., Axon D.~J., Chiaberge M., Macchetto F.~D., Sparks W.~B., 2010, A\&A, 509, A260000 

\bibitem[\protect\citeauthoryear{Buttiglione et 
al.}{2009}]{2009A&A...495.1033B} Buttiglione S., Capetti A., Celotti A., Axon D.~J., Chiaberge M., Macchetto F.~D., Sparks W.~B., 2009, A\&A, 495, 1033 

\bibitem[\protect\citeauthoryear{Chabrier}{2003}]{2003PASP..115..763C} 
Chabrier G., 2003, PASP, 115, 763 


\bibitem[\protect\citeauthoryear{Cid Fernandes et 
al.}{2005}]{2005MNRAS.358..363C} Cid Fernandes R., Mateus A., Sodr{\'e} L., 
Stasi{\'n}ska G., Gomes J.~M., 2005, MNRAS, 358, 363 

\bibitem[\protect\citeauthoryear{Cid Fernandes et 
al.}{2009}]{2009RMxAC..35..127C} Cid Fernandes R., et al., 2009, RMxAC, 35, 
127

\bibitem[\protect\citeauthoryear{Cid Fernandes et 
al.}{2010}]{2010MNRAS.403.1036C} Cid Fernandes R., Stasi{\'n}ska G., 
Schlickmann M.~S., Mateus A., Vale Asari N., Schoenell W., Sodr{\'e} L., 
2010, MNRAS, 403, 1036 



\bibitem[\protect\citeauthoryear{Condon et al.}{1998}]{1998AJ....115.1693C} 
Condon J.~J., Cotton W.~D., Greisen E.~W., Yin Q.~F., Perley R.~A., Taylor 
G.~B., Broderick J.~J., 1998, AJ, 115, 1693

\bibitem[\protect\citeauthoryear{Edge et al.}{1959}]{1959MmRAS..68...37E} 
Edge D.~O., Shakeshaft J.~R., McAdam W.~B., Baldwin J.~E., Archer S., 1959, 
MmRAS, 68, 37 

\bibitem[\protect\citeauthoryear{Fanaroff 
\& Riley}{1974}]{1974MNRAS.167P..31F} Fanaroff B.~L., Riley J.~M., 1974, MNRAS, 167, 31P 

\bibitem[\protect\citeauthoryear{Fitzpatrick}{1999}]{1999PASP..111...63F} 
Fitzpatrick E.~L., 1999, PASP, 111, 63 

\bibitem[\protect\citeauthoryear{Gower, Scott, 
\& Wills}{1967}]{1967MmRAS..71...49G} Gower J.~F.~R., Scott P.~F., Wills D., 1967, MmRAS, 71, 49

\bibitem[\protect\citeauthoryear{Hales, Baldwin, 
\& Warner}{1993}]{1993MNRAS.263...25H} Hales S.~E.~G., Baldwin J.~E., Warner P.~J., 1993, MNRAS, 263, 25 

\bibitem[\protect\citeauthoryear{Hales, Baldwin, 
\& Warner}{1988}]{1988MNRAS.234..919H} Hales S.~E.~G., Baldwin J.~E., Warner P.~J., 1988, MNRAS, 234, 919 

\bibitem[\protect\citeauthoryear{Hales et al.}{1990}]{1990MNRAS.246..256H} 
Hales S.~E.~G., Masson C.~R., Warner P.~J., Baldwin J.~E., 1990, MNRAS, 
246, 256 

\bibitem[\protect\citeauthoryear{Hales et al.}{1993}]{1993MNRAS.262.1057H} 
Hales S.~E.~G., Masson C.~R., Warner P.~J., Baldwin J.~E., Green D.~A., 
1993, MNRAS, 262, 1057 

\bibitem[\protect\citeauthoryear{Hales et al.}{1991}]{1991MNRAS.251...46H} 
Hales S.~E.~G., Mayer C.~J., Warner P.~J., Baldwin J.~E., 1991, MNRAS, 251, 
46 

\bibitem[\protect\citeauthoryear{Hales et al.}{2007}]{2007MNRAS.382.1639H} 
Hales S.~E.~G., Riley J.~M., Waldram E.~M., Warner P.~J., Baldwin J.~E., 
2007, MNRAS, 382, 1639 

\bibitem[\protect\citeauthoryear{Hales et al.}{1995}]{1995MNRAS.274..447H} 
Hales S.~E.~G., Waldram E.~M., Rees N., Warner P.~J., 1995, MNRAS, 274, 447 

\bibitem[\protect\citeauthoryear{Hardcastle et 
al.}{1998}]{1998MNRAS.296..445H} Hardcastle M.~J., Alexander P., Pooley 
G.~G., Riley J.~M., 1998, MNRAS, 296, 445 


\bibitem[\protect\citeauthoryear{Hine 
\& Longair}{1979}]{1979MNRAS.188..111H} Hine R.~G., Longair M.~S., 1979, MNRAS, 188, 111 

\bibitem[\protect\citeauthoryear{Janda}{2006}]{Janda2006}
Janda, K. 2006, MSc Dissertation, Jagiellonian University, Krakow

\bibitem[\protect\citeauthoryear{Kaiser, Schoenmakers, {\ 
R&ouml}ttgering}{2000}]{2000MNRAS.315..381K} Kaiser C.~R., Schoenmakers A.~P., R{\"o}ttgering H.~J.~A., 2000, MNRAS, 315, 381 

\bibitem[\protect\citeauthoryear{Kauffmann, Heckman, 
\& Best}{2008}]{2008MNRAS.384..953K} Kauffmann G., Heckman T.~M., Best P.~N., 2008, MNRAS, 384, 953

\bibitem[\protect\citeauthoryear{Kauffmann 
\& Heckman}{2009}]{2009MNRAS.397..135K} Kauffmann G., Heckman T.~M., 2009, MNRAS, 397, 135 


\bibitem[\protect\citeauthoryear{Le Borgne et 
al.}{2003}]{2003A&A...402..433L} Le Borgne J.-F., et al., 2003, A\&A, 402, 433

\bibitem[\protect\citeauthoryear{Ledlow 
\& Owen}{1996}]{1996AJ....112....9L} Ledlow M.~J., Owen F.~N., 1996, AJ, 112, 9 

\bibitem[\protect\citeauthoryear{Lin et al.}{2010}]{2010ApJ...723.1119L} 
Lin Y.-T., Shen Y., Strauss M.~A., Richards G.~T., Lunnan R., 2010, ApJ, 
723, 1119

\bibitem[\protect\citeauthoryear{Liu et al.}{2010}]{2010ApJ...715L..30L} 
Liu X., Greene J.~E., Shen Y., Strauss M.~A., 2010a, ApJ, 715, L30 

\bibitem[\protect\citeauthoryear{Liu et al.}{2010}]{2010ApJ...708..427L} 
Liu X., Shen Y., Strauss M.~A., Greene J.~E., 2010b, ApJ, 708, 427 

\bibitem[\protect\citeauthoryear{Machalski, Koziel-Wierzbowska, 
\& Jamrozy}{2007}]{2007AcA....57..227M} Machalski J., Koziel-Wierzbowska D., Jamrozy M., 2007, AcA, 57, 227 

\bibitem[\protect\citeauthoryear{Marecki 
\& Swoboda}{2011}]{2011A&A...525A...6M} Marecki A., Swoboda B., 2011, A\&A, 525, A6 

\bibitem[\protect\citeauthoryear{Mateus et al.}{2006}]{2006MNRAS.370..721M} 
Mateus A., Sodr{\'e} L., Cid Fernandes R., Stasi{\'n}ska G., Schoenell W., 
Gomes J.~M., 2006, MNRAS, 370, 721 

\bibitem[\protect\citeauthoryear{Owen 
\& Laing}{1989}]{1989MNRAS.238..357O} Owen F.~N., Laing R.~A., 1989, MNRAS, 238, 357

\bibitem[\protect\citeauthoryear{Owen 
\& Ledlow}{1994}]{1994ASPC...54..319O} Owen F.~N., Ledlow M.~J., 1994, ASPC, 54, 319 

\bibitem[\protect\citeauthoryear{Pearson}{1975}]{1975MNRAS.171..475P} 
Pearson T.~J., 1975, MNRAS, 171, 475 

\bibitem[\protect\citeauthoryear{Pearson 
\& Kus}{1978}]{1978MNRAS.182..273P} Pearson T.~J., Kus A.~J., 1978, MNRAS, 182, 273 

\bibitem[\protect\citeauthoryear{Pierce et al.}{2010}]{2010MNRAS.405..718P} 
Pierce C.~M., et al., 2010, MNRAS, 405, 718

\bibitem[\protect\citeauthoryear{Pilkington 
\& Scott}{1965}]{1965MmRAS..69..183P} Pilkington J.~D.~H., Scott P.~F., 1965, MmRAS, 69, 183 

\bibitem[\protect\citeauthoryear{Rawlings et 
al.}{1989}]{1989MNRAS.240..701R} Rawlings S., Saunders R., Eales S.~A., 
Mackay C.~D., 1989, MNRAS, 240, 701 

\bibitem[\protect\citeauthoryear{Rees}{1990}]{1990MNRAS.244..233R} Rees N., 
1990, MNRAS, 244, 233 

\bibitem[\protect\citeauthoryear{Schinnerer et 
al.}{2007}]{2007ApJS..172...46S} Schinnerer E., et al., 2007, ApJS, 172, 46 

\bibitem[\protect\citeauthoryear{Smol{\v c}i{\'c} et 
al.}{2009}]{2009ApJ...696...24S} Smol{\v c}i{\'c} V., et al., 2009, ApJ, 
696, 24 

\bibitem[\protect\citeauthoryear{Smol{\v 
c}i{\'c}}{2009}]{2009ApJ...699L..43S} Smol{\v c}i{\'c} V., 2009, ApJ, 699, 
L43 

\bibitem[\protect\citeauthoryear{Spergel et 
al.}{2003}]{2003ApJS..148..175S} Spergel D.~N., et al., 2003, ApJS, 148, 
175 

\bibitem[\protect\citeauthoryear{Stasi{\'n}ska et 
al.}{2006}]{2006MNRAS.371..972S} Stasi{\'n}ska G., Cid Fernandes R., Mateus 
A., Sodr{\'e} L., Asari N.~V., 2006, MNRAS, 371, 972 (S06)

\bibitem[\protect\citeauthoryear{Tremaine et 
al.}{2002}]{2002ApJ...574..740T} Tremaine S., et al., 2002, ApJ, 574, 740 

\bibitem[\protect\citeauthoryear{Waldram et 
al.}{2003}]{2003MNRAS.342..915W} Waldram E.~M., Pooley G.~G., Grainge 
K.~J.~B., Jones M.~E., Saunders R.~D.~E., Scott P.~F., Taylor A.~C., 2003, 
MNRAS, 342, 915 

\bibitem[\protect\citeauthoryear{York et al.}{2000}]{2000AJ....120.1579Y} 
York D.~G., et al., 2000, AJ, 120, 1579 

\bibitem[\protect\citeauthoryear{Zirbel 
\& Baum}{1995}]{1995ApJ...448..521Z} Zirbel E.~L., Baum S.~A., 1995, ApJ, 448, 521 

\end{thebibliography}
\end{document}